\documentclass[a4, 12pt]{article}
\usepackage{setspace}
\usepackage{graphicx,amsmath,amsfonts,amssymb,bm, url, color, dsfont}
\usepackage{fullpage}
\usepackage[small,bf]{caption}
\usepackage{authblk}
\usepackage{hyperref}
\hypersetup{
    colorlinks=true,%
    citecolor=blue,%
    filecolor=blue,%
    linkcolor=blue,%
    urlcolor=blue
}

\begin{document}


\title{\textbf{\textsf{Afferent specificity, feature specific connectivity influence orientation selectivity: A computational study in mouse primary visual cortex}}}
\author[*]{Dipanjan Roy}
\author[*]{Yenni Tjandra}
\author[*]{Konstantin Mergenthaler}
\author[**]{Jeremy Petravicz}
\author[**]{Caroline A. Runyan}
\author[**]{Nathan R. Wilson}
\author[**]{Mriganka Sur}
\author[*]{Klaus Obermayer}
\affil[*]{Department of Theoretical Computer Science, Technische Universitat Berlin}
\affil[**]{Picower Institute for Learning and Memory, Department of Brain and Cognitive Sciences, Massachusetts Insitute of Technology,
Cambridge, MA.}
\renewcommand*\abstractname{\textsf{Abstract}}

\newcommand{\bs}{\boldsymbol}
\newcommand{\smum}{\, \mu \mathrm{m}}
\newcommand{\mums}{\, \mu \mathrm{m}/\mathrm{s}}

\maketitle


\abstract{Primary visual cortex (V1) provides crucial insights
into the selectivity and emergence of specific output features such as orientation tuning. Tuning and selectivity of cortical neurons in mouse visual cortex is not equivocally resolved so far.
While many in-vivo experimental studies found inhibitory neurons of all subtypes to be broadly tuned for orientation other studies report 
inhibitory neurons that are as sharply tuned as excitatory neurons. 
These diverging findings about the selectivity of excitatory and inhibitory cortical neurons prompted us to ask the following questions: (1) How different or similar is the cortical computation with that in previously described species that relies on map? (2) What is the network mechanism underlying the sharpening of orientation selectivity in the mouse primary visual cortex?  Here, we investigate the above questions in a computational framework with a recurrent network composed of Hodgkin-Huxley (HH) point neurons. Our cortical network with random connectivity alone could not account for all the experimental observations, which led us to hypothesize, (a) Orientation dependent connectivity (b) Feedforward afferent specificity to understand orientation selectivity of V1 neurons in mouse. Using population (orientation selectivity index) OSI as a measure of neuronal selectivity to stimulus orientation we test each hypothesis separately and in combination against experimental data. Based on our analysis of orientation selectivity (OS) data we find a good fit of network parameters in a model based on afferent specificity and connectivity that scales with feature similarity. We conclude that this particular model class best supports datasets of orientation selectivity of excitatory and inhibitory neurons in layer 2/3 of primary visual cortex of mouse.} 
\section*{Introduction}
Neuronal response selectivity for stimulus orientations is one of the key properties that mouse V1 shares with other species~\cite{Dräger_1975, VanHooser2006, Niell2008, VandenBergh2010, Tan2011}. In higher mammals, anatomically close V1 neurons have similar preferred orientations (PO), giving rise to orientation maps. Orientation maps in turn correlate with the connectivity that is  observed in V1 among all the cell types. The fine-scale organization of mouse cortical microcircuits is clearly dissimilar~\cite{KerlinReid_2010, Jia_2010, Smith2010}. In particular, receptive fields of layer 2/3 neurons were found to be relatively large with high overlap for neighboring neurons~\cite{Smith2010}. In addition, a salt-and-pepper organization of orientation preference exists in layer 2/3 \cite{Lee_Reid_2011, Sohya2007a}. Broad tuning of V1 cell types is a direct consequence of such salt-and-pepper organization\cite{Liu2009, Liu_2011, KerlinReid_2010}. Cortical mechanism for tuning and selectivity of cat and monkey V1 neurons have been extensively investigated with different category of computational models so far \cite{Priebe_Ferster_2008, Shapley_Hawken_Ringach_2003, Jia_2010, Tan2011, Sompolinsky_Shapley_1997, Marino2005}.  In comparison, there are very few model based analysis in mouse cortex, lone, a recent model  
from Hansel et. al. that sheds light on the orientation selectivity of V1 neurons without a functional map\cite{Hansel2012}.  In  a previous work,  it was established that the experimentally measured response properties of cat V1  neurons are consistent with the predictions of a Hodgkin-Huxley
network model dominated by recurrent interactions and with balanced contributions
from excitation and inhibition \cite{Marino2005}. However, this cannot rule out
alternative cortical operating regimes.  Subsequently, it has been shown in a recent study, with model-based  data analysis in V1 of cat;  that close to pinwheel centers recurrent inputs show variable orientation preferences; within iso-orientation domains inputs are relatively uniformly tuned. Physiological properties such as cells membrane potentials, spike outputs, and conductances change systematically with map location. With this background in mind, 
we would like to turn to model-based data analysis in order to assess a continuum of network models, that encompasses the full range from feed-forward via inhibition- and- excitation (I-E) dominated models to models with excitation and inhibition in balance. Thus, in contrast to many modeling approaches, we try to determine the space of models able to account for the data, rather than demonstrating one model to be compatible with the data sets.  Our computational model is composed of HH point neurons with realistic AMPA, NMDA, GABA synapses and a salt-pepper organization of V1 neurons. Model assumptions are (a) random lateral connectivity, (b) Orientation dependent connectivity, (c)  Feed-forward afferent specificity, (d) combining afferent specificity and orientation dependent connectivity.
 Our experimental database comprise of OSI, OI values of membrane potential, spike output, conductances  of V1 neurons  as 
published by several recent studies. More specifically, we are using OSI and OI values for all four responses (spike, membrane potential, excitatory and inhibitory conductance) from in vivo whole-cell voltage-clamp recordings from excitatory cells in layer 2/3 of mouse visual cortex~\cite{Liu2009,Tan2011}.
 Also, we have used the firing rate excitatory and inhibitory population OSI reported by~\cite{Niell2008, Runyan2010}. 
 We carry out a goodness of fit test using K-S test between simulated and experimental OSI distributions. We conclude the test results  
as a good fit if p values are above 0.05 threshold which defines $95\%$ significance level. Based on the test results, we conclude that the likely models to account for the experimental observations are the models with afferent specificity and afferent specificity in conjunction with feature specific lateral connectivity. 
\section{Materials and Methods}

{\itshape Network architecture}
Over all our network model is composed of three neuron populations: Excitatory
afferent neurons provide feed forward input to the primary visual cortex
layer, which is received by an excitatory and an inhibitory neuron populatio (see: Fig.~\ref{fig:mapModel}). Neurons are
spread on a two-dimensional $D \times D$ grid representing a square patch
of mouse V1 of approximately 0.5 x 0.5 mm$^2$. Every grid point is occupied by
an excitatory neuron (population size: $N_E = D^2$), but only every third
randomly-selected site is occupied by an inhibitory neuron (population size: $N_I =
N_E/3$) is placed. Neurons at every site are highly interconnected and receive
input from a number of feed-forward neurons
$N_A$ with stimulus specific rates.

{\itshape Primary visual cortex neurons}
Each visual cortex cell is describe by a Hodgkin-Huxley-type neuron. It's membrane
potential dynamic $V$ is driven by several intrinsic and synaptic currents:
\begin{equation}
\label{MPdynamic}
C_m \frac{dV}{dt}=-g_{L}(V-E_L)-I_{int}-I_{syn}-I_{bg},
\end{equation}
where $C_m$ is the membrane capacitance, $t$ the time, $g_{L}$ the leak
conductance, $E_L$  reversal potential, and $I_{int}$,
$I_{syn}$,  $I_{bg}$ are voltage dependent intrinsic, synaptic, and
background currents, respectively. To account for different membrane properties of
excitatory and inhibitory neurons the leak conductance $g_{L}$ is selectively specified as $g_{EL}$ and $g_{IL}$
for excitatory and inhibitory neurons. Explicit parameterizations of the model are comprised in Table~\ref{tab:HHModel}
{\itshape Intrinsic Currents, gating variables, and ion channel properties}
Three intrinsic currents, a fast $Na^+$ current $I_{Na}$, a delayed-rectified $K^+$ current 
$I_{Kd}$ and a slow non-inactivating $K^+$ current $I_M$ are summed to comprise
 the total intrinsic current to each neuron. Intrinsic currents are based on
 voltage dependent gating variables:
\begin{equation}
I_Y (t) = \bar g_Y m_{Yact}^{l_Y}(t) m_{Yinac}^{k_Y}(t) (V(t) - E_Y)
\end{equation}
where $\bar g_Y$ is the peak conductance, $E_Y$ is the reversal
potential, $l_Y$,$k_Y$ number of activation and inactivation sites, and $m_{Yact}$, $m_{Yinac}$ are activation and inactivation dynamics,
with $Y \in \{Na,Kd,M\}$. A differentiation of $ \bar g_M$ for
excitatory $\bar g_{EM}$ and inhibitory $\bar g_{IM}$ accounts for reduced
adaptation of inhibitory neurons. Activation and inactivation dynamics are described by:
\begin{eqnarray*}
\frac{dm_{Y}}{dt} &=& t_Y \alpha_{Y1} ( 1 - m_{Y}) - \alpha_{Y2} m_{Y} \\
 \alpha_{Yi} &=& \frac{c_1 c_2(V)}{\exp(c_3(V)) + c_4}
\end{eqnarray*} 
For the particular receptor kinetics values from Destexhe \& Pare
\cite{Destexhe1999} are used (see: Table~\ref{tab:receptor})
$\emph{Synaptic background current}$
For embedding the small simulated network into a larger surrounding cortex each neuron
receives an unspecific fluctuating synaptic background current $I_{bg}$
mediated by excitatory ($g_{bgE}$) and inhibitory ($g_{bgI}$) conductances: 
\begin{equation}
I_{bg}=g_{bgE}(V(t)-E_{bgE})-g_{bgI}(V(t)-E_{bgE})
\end{equation}
which follow a stochastic process similar to an Ornstein-Uhlenbeck
process~\cite{Destexhe_2003}. $g_{bgX}$ is given by:
\begin{equation}
g_{bgX}(t+d t)= g^0_{bgX} + [g_{bgX}(t)-g^0_{bgX}]exp(-d
t/\tau_{bgX})+\sqrt{1-\exp(-2 d t/ \tau_{bgX})}N(0,\sigma_{bgX})
\end{equation} 
where $g^0_{bgX}$ is the average conductance, $\tau_{bgX}$ is the background synaptic time constant, and $N(0,\sigma_{bgX})$ is a normally
distributed random number with zero mean and standard deviation
$\sigma_{bgX}$ and $X \in \{E,I\}$.

{\itshape Synaptic currents and receptor dynamics}
Synaptic currents to each neuron originate via lateral connections from
neighboring excitatory as well as inhibitory neurons and via feed-forward
connections from excitatory afferent. The total synaptic current is given by\\
\begin{equation}
I_{syn}(t) = \left[ \underbrace{\frac{ \bar{g}_A }{N_A (t)} g_A
    (t)}_{\textnormal{afferent\:excitatory}} + \underbrace{\frac {\bar{g}_E}{N_E}\left(q g_{E1}
  (t) + (1-q) g_{E2}(t)\right) }_{\textnormal{recurrent\:excitatory}}\right]
(V - E_e) - \underbrace{\frac{\bar{g}_I}{N_I} g_I(t) (V -
  Ei)}_{\textnormal{recurrent\:inhibitory}}
\end{equation}

where all $\bar g$,$N$, and $g(t)$ are peak conductance, number of
connections, and time dependent conductance for the specific type of
connection, $q$ quantifies the fraction of fast AMPA-like receptors and $E_e$
and $E_i$ are the synaptic reversal potentials. Time dependent conductances for fast excitatory
$g_A$,$g_{E1}$, and fast inhibitory GABA$_A$ type $g_{I}$ synapses are
described by a delayed (by $\Delta t$) exponential response to spikes at time $t_{sp}$: 
\begin{equation}
g_X(t)=\frac{1}{\tau_X}\sum\limits_{t_k<t} e^{\frac{t_k -t}{\tau_X}}  
\end{equation}
with $t_k = t_{sp} + \Delta t$ and $\tau_X$ with $X \in \{A,E1,I\}$ is the
synaptic time-constant. Multiple events are linearly summed. For each
pre-synaptic neuron and used to all postsynaptic neurons one delay value
$\Delta t$, which comprise synaptic and conductance delays and is drawn from a truncated Gaussian
($\max(1/\sqrt{2\pi\sigma_{X\Delta}}
\exp(-\mu_{X\Delta}^2/2\sigma_{X\Delta}^2),d t)$), with $X \in \{E,I\}$
and $d t$ the numeric integration time-step.
 For the slower response kinetic of
NMDA-type receptors a bi-exponetial description is applied:
\begin{equation}
g_{E2}(t)=\frac{1}{\tau_f-\tau_r}\sum\limits_{t_k<t}\left( e^{\frac{t_k
    -t}{\tau_f}} - e^{\frac{t_k - t}{\tau_r}} \right) 
\end{equation}
where $\tau_f$ and $\tau_r$ are rise and decay time constants and again multiple events are linearly summed. Both types of conductances are normalized
so that for a single incoming spike, the integral over time is 1.
{\itshape Lateral connectivity}

Each neuron receives recurrent synaptic input from neighboring excitatory and
inhibitory neurons. Each excitatory (and inhibitory) neuron gets a fixed
number of  incoming recurrent connections from excitatory $N_{EE}$ ($N_{IE}$)
and inhibitory neurons $N_{EI}$ ($N_{II}$). Connections are established by
drawing randomly from a 2-d radial-symmetric Gaussian probability distribution
(see Figure~\ref{fig:mapModel}:
\begin{equation}
P_X(x)=\left\{ \begin{array}{ll}
 0 & \mbox{for $x=0$ (no self-connections)};\\
  1/\sqrt{2\pi\sigma_X}\exp(-x^2/{2\sigma_X^2}) & \mbox{otherwise,}.\end{array} \right.
\end{equation}
where $x$ is the distance to the presynaptic neuron (in pixels) and
$\sigma_X$ with $X \in \{E,I\}$ the width of the Gaussian. Connections at the
boundaries are build using periodic boundary conditions.
{\itshape Salt-pepper architecture of rodent V1}

The ``salt-and-pepper''-organization of preferred orientation in rodent
primary visual cortex is implemented by uncorrelated random assignment of preferred
orientation to afferent neurons projecting to one site in the visual cortex
lattice (to the excitatory as well as inhibitory neuron at the same site). As
each afferent neuron is only projecting to one site and all afferent neuron
projecting to one site have the same preferred orientation primary visual
cortex neurons inherit preferred orientation from their afferent. Each
afferent neuron is realized as a Poisson process with a stimulus specific
firing rate:
 \begin{equation}
f_{Aff}(t,\theta_{stim})= f_{Amax}\cdot \frac{\log(C+1)}{\log(101)}\cdot(r_{Aff}(\theta_{stim})+r_{base}),
\end{equation}
\begin{equation}
r_{Aff}(\theta_{stim})=(1-r_{base})\cdot\exp\bigl(-\frac{(\theta_{stim}-\theta)^2}{(2\sigma_{Aff})^2}\bigr),
\end{equation} 
where $\theta_{stim}$ is the orientation of the presented stimulus, $f_{Amax}$
is the maximal firing rate when optimally stimulated, $\theta$ is the preferred orientation chosen according to the neuron's location in the artificial orientation map, $\sigma_{Aff}$ is the
orientation tuning width, and $r_{base}$ is a baseline response without stimulus.
$C$ is the stimulus contrast in $\%$ (100 \% for all simulations carried out here).
$\emph{Numerical procedures and analysis of simulation results}$

The network model was implemented in Python\texttrademark  2, using simulator for spiking neural network Brian, numerical 
and scientific computing package NumPy and SciPy. The simulations were performed for 1.2 s with a fixed time step of 
$dt=0.05 ms$. The first 200 ms of simulation aimed to set the network to a steady state. Thus, the results within 
this time were ignored. The membrane potential ($Vm$), the firing rate ($f$), the excitatory and inhibitory conductance
($g_e$ and $g_i$) are captured for each cells for the last 1 s. The mean membrane potential was calculated by omitting 
the values from 4 ms before until 7 ms after every spike. The mean of this value then was subtracted with the resting potential 
of the cell ($V_m=-67.5$mV) for calculating the sharpness of tuning. The total excitatory conductance $g_e$ is the sum of the 
afferent conductance, the recurrent of excitatory conductance of the fast AMPA-like and slow NMDA-like excitatory synapses, 
and the excitatory background conductance. The total inhibitory conductance $g_i$ is the sum of the recurrent inhibitory 
conductance of GABA$_A$-like inhibitory synapse, the conductance of the non-inactivating K$^+$ current and the inhibitory background conductance.

\emph{Parameter space explorations and model modifications} 

In Hodgkin-Huxley network model V1 neurons receive both feedforward afferent excitatory input
which are moderately tuned for orientation and as well as, recurrent input from neighboring V1 
neurons. Four types of recurrent connections are implemented in V1 network namely E to E, I to E, E to I and I to I respectively. Afferent input and lateral connectivity are the two key components of this network model. 
To implement it we assume afferent excitation that each V1 cell (both inhibitory and excitatory) receives comprise of $N_{Aff}$  excitatory neurons, which are sufficiently well tuned to orientation in the default network (where afferent specificity and feature specific lateral connectivity is not introduced). 
The operating regimes of a ﬁring rate model can be quantified based  on the strength and shape of the effective recurrent input \cite{Kang_2003}. Results proposed by Kang et al. \cite{Kang_2003}, however, are based on the analytical solution of a linear ﬁring rate model where all neurons are above threshold and cannot be extended to the non-linear Hodgkin-Huxley network model used here.
Hence, we resort to numerics in our present study and systematically vary the strength of recurrent excitation and inhibition (physiological network parameters) relative to the strength of our afferent input to V1 cells in order to characterize different operating regimes. We carry out a full network simulation with default set of network parameters with 19 different maximum conductance values of excitatory connection to excitatory $(\bar g_{EE})$ and 14 different maximum conductance values of excitatory connection to inhibitory neurons $(\bar g_{IE})$, while keeping all other parameters fixed. 
The maximum conductance values are ranging from 0.1 to 1.9 with a step size of 0.1 for $\bar g_{EE}$ and from 0.033 to 2.733 with a step size of 0.2 for  $\bar g_{IE}$. These values are multiples of the maximum afferent conductance of excitatory neuron $\bar g_{E}^{Aff}$. The excitatory and inhibitory recurrent and afferent input currents were computed for each cell in all parameter combinations. Both recurrent input currents were normalized with afferent input current at preferred orientation. As described in \cite{Martin2010}, the operating regimes for Hodgkin-Huxley (HH) network model can be numerically classified using following criteria: 
feedforward (FF), if the sum of absolute excitatory and inhibitory currents are below 0.3; excitatory (EXC), if the excitatory current dominates the inhibitory current for all presented orientations and 
the sum of absolute excitatory and inhibitory currents are above 0.3; 
inhibitory (INH) if the inhibitory current dominates the excitatory current and the sum of excitatory, inhibitory currents are below 0; reccurent (REC), if the excitatory and inhibitory currents are strong and balanced and their sum is approximately 0.
 
\emph{Tuning of afferent and their specificity to V1 neurons} 
By default in our "salt-and-pepper" network model each afferent neuron is realized as a Poisson process with a stimulus specific firing rate calculated based on Eq.(9) and (10). Afferent tuning curves are all based on a Gaussian in Eq.(10) where the orientation tuning width $\sigma_{Aff}$ small. Hence, in this model these excitatory afferent neurons projecting to one site in the visual cortex lattice (to the excitatory as well as inhibitory neuron at the same site) are sufficiently well tuned. One of the justification for such selection is that the recurrency either amplifies or dampens output spike response of cortical cells modestly. This we have tested using different afferent tuning width (from small to large). Based on our results we have then decided to held 
$\sigma_{Aff}$ = 27.5$^o$ without the loss of any generality. 

\emph{Afferent speficity to V1 neurons}

Afferent specificity is implemented by adding an untuned afferent excitatory input component 
to receiving V1 inhibitory cells. In this case, afferent excitatory firing rates are computed 
using a truncated Gaussian as follows,
Let $X=(X_{1},\ldots,X_{d})$ be a $d-$dimensional Gaussian vector with
mean $\mu$ and standard deviation $\sigma$, and let $[a_{i},b_{i}]$
be $d$ intervals, where $b_{i}$ may be either a real number or $+\infty$. 
The distribution of $X$, conditional on the event that $X_{i}\in[a_{i},b_{i}]$,
$i=1,\ldots,d$, is usually called a truncated Gaussian distribution.
First, we consider the problem of simulating a random variable $\theta$ from
a univariate Gaussian density truncated to $[\theta_{a},\theta_{b})$: 
\begin{equation}
p(\theta)=\frac{\varphi(\theta)}{\Phi(-\theta_{a})}I(x\geq \theta_{a})\label{eq:tn}
\end{equation}
Hence, for some truncation point $\theta_{a}$, where $\varphi$ and  $\Phi$ denote respectively the unit Gaussian probability density and cumulative distribution functions.
Now, based on Eq.(11) we compute $p(\theta)$ and substitute in Eq.(10) 
for firing rate at a preferred orientation of our stimulus $\theta_{stim}$.
In order to constrain choice of $\mu$ and $\sigma$ in our formulation we 
carry out a grid search over this 2D parameter space and subsequently compute 
$p$ values based on Kolmogorov-Smirnov goodness of fit test to arrive at optimal values. The combination of $\mu$ and $\sigma$ are then selected based on the computed $p$ values to carry out further parameter grid search over recurrent excitation and inhibition values (network parameters used in each model space). 
Afferent excitatory input is further splitted into 2 components, (a) $N_{t}$tuned, (b)$N_{u}$ untuned, which is received by a layer of V1 inhibitory neurons. Fraction of tuned vs untuned components are decided again based on a grid search. They are held fixed throughout the afferent specific simulations are concerned. 

$\emph{Measures of selectivity, analysis of simulated data and fit}$

The orientation selectivity were measured with Orientation Selectivity Index (OSI), Orientation Modulation Index (OMI) and Orientation Index (OI). The OSI is given by 
\cite{Swindale1998}:\\
\begin{equation}
\label{OSI}
OSI=\frac{\sqrt{(\sum\limits_i R(\theta_i)\cos(2\theta_i))^2+(\sum\limits_i{R(\theta_i)\sin(2\theta_i)})^2}}{\sum\limits_i{R(\theta_i)}}
\end{equation}
where $\theta$ is the angle of stimulus ranging from -90$^o$ to 90$^o$, $R(\theta)$ is the quantity of response as the stimulus
was presented. The value of OSI is ranging from 0 to 1. If the response is only significant to one stimulus orientation 
(its preferred orientation), the OSI will have value 1. Whereas, if the responses are equal to all presented orientation, 
the OSI is equal 0. 
In this study, the OSI quantified the orientation selectivity of the mean membrane potential ($Vm$), the firing rate ($f$), 
the mean excitatory and inhibitory conductances ($g_e$ and $g_i$). The background conductance were excluded from both mean 
conductance for calculating the OSI. 
The OSI is also used to determine orientation selectivity of recurrent input received by a neuron based 
on its position on the map, which is referred as map OSI. This map OSI is calculated by binning the orientation preference 
of all pixels within a fixed radius around a neuron into bins of 10$^o$ size. The number of cells in each bin replaces the 
quantity of response $R(\theta)$. This study used a radius size of 10 pixels , instead of 8 pixels( $\approx$ 250 $\mu m$), 
which is described by \cite{Stimberg2009}. The choice of 10 pixels was based on the saturation of the intercept of OSI-OSI regression line (see for further details \cite{Stimberg2009}) (data not shown). 

The experimental OSI often defined as in\cite{Niell2008} 
\begin{equation}
OSI=\frac{R_{pref}-R_{ortho}}{R_{pref}+R_{ortho}},
\end{equation}
where, $R_{pref}$ is the peak response in the preferred orientation, and $R_{ortho}$
is the response in the orthogonal direction. Orientation selectivity index (OSI) representing 
the tuned versus untuned component of the response. 
We rename the above measure here as OMI in order to avoid any confusion with OSI already introduced in Eq.\ref{OSI}

The OI as defined by \cite{Tan2011}: 
\begin{equation}
OI=1-\frac{N}{P}
\end{equation}  
where N and P are the quantity of response to null (orthogonal) orientation and the preferred orientation, respectively. 
The OI is used to quantify the tuning of the mean excitatory ($g_e$) and inhibitory ($g_i$) conductance. The background conductance are included in both mean conductance.

$\emph{Goodness of fit and nonparametric test}$ 

A two-sample Kolmogorov-Smirnov (K-S) test is to compare the distributions of the values in the two data vectors. The null hypothesis is that compared datavectors are from the same continuous distributions. The alternative hypothesis is that they are from different continuous distributions. Statistics aim to assign numbers to the test results; p-values report if the numbers differ significantly within $p \le 0.05$ at the $5\%$ significance level and $p \le 0.001$ at the $1\%$ significance level. We reject the null hypothesis if $p$ is "smaller" than this number. 

\section{Results}
\textbf{Experimental data on orientaion selectivity are nonoverlapping and mixed}

\begin{figure}[h!]
    \includegraphics[scale=0.32]{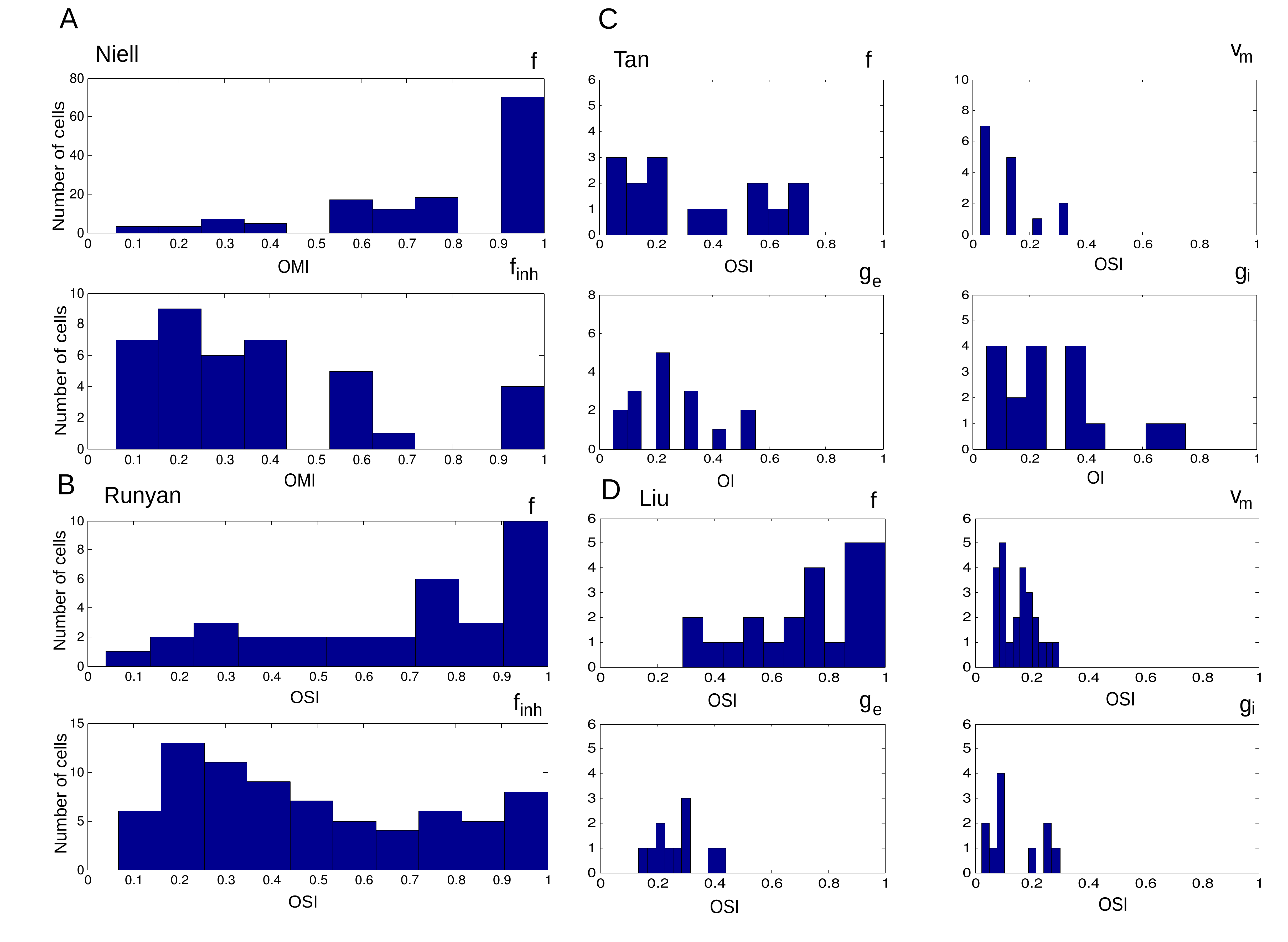}
     \caption[Experimental results]{Histograms of OSI of spikes, membrane potential,
    inhibitory and excitatory conductances are shown here respectively for various intercellular recordings from the superficial depth of the visual cortex of mouse. 
    Figure (A) shows the firing rate OSI distribution of excitatory (In the top row) and inhibitory cells (in the bottom row), data published by Niell et. al.~\cite{Niell2008} and figure (B) show the spikes recorded from inhibitory neurons (total number of cells, n = 74) and excitatory neurons (total number of cells, n = 34) published by Runyan et. al.\cite{Runyan2010}.
The orientation selectivity of PV+ inhibitory neurons shows a multimodal distribution. 
Because of a large group of untuned inhibitory neurons, the mean orientation selectivity index (OSI) of inhibitory population is lower than that of excitatory population (p $<$ 0.5); however, a second mode centered around OSI = 0.8-1.0 in the inhibitory population distribution suggests a second subtype of inhibitory neurons with high selectivity. 
Figure (C) shows the spike, membrane potential OSI (top row) for all excitatory cells and OI distributions for conductances for excitatory cells (bottom row) published by Tan et. al.~\cite{Tan2011}. 
Figure (D) shows the spike, membrane potential OSI distributions (top row) and the inhibitory, excitatory conductance OSI distributions (bottom row)  published by Liu et. al.~\cite{Liu_2011}} 
\label{fig:OSIExp}
\end{figure}

\begin{figure}[hc]
  \includegraphics[scale=0.36]{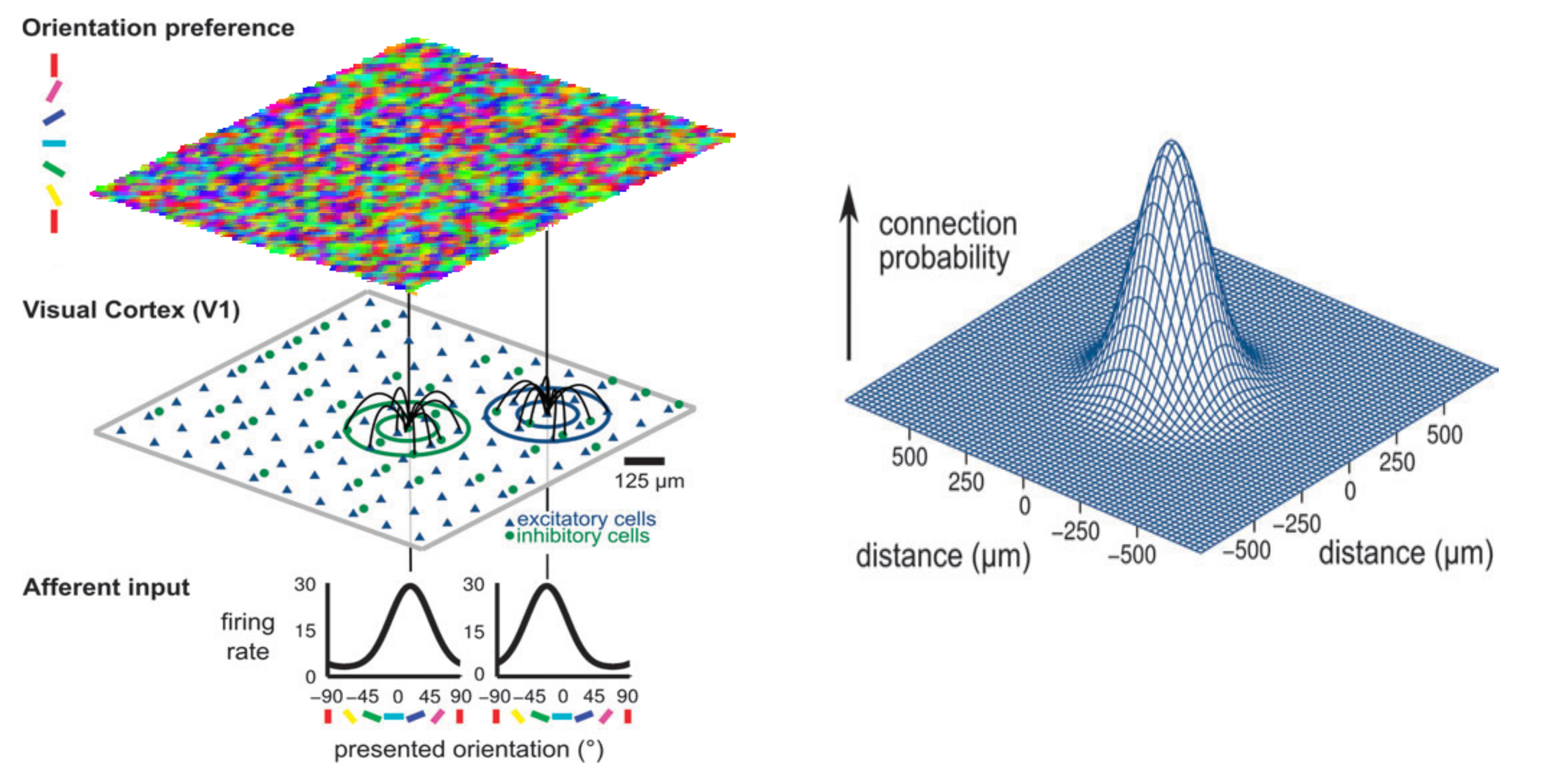} 
\caption[A salt-pepper map model]{Salt-pepper architecture of rodent V1. The cartoon sketches 
the architecture of network model classes implemented here in a computational framework: A layer of excitatory 
(blue triangles) and inhibitory neurons (green circles) receives afferent as well as lateral input. In the Hodgkin-Huxley 
networks, the number of inhibitory cells is one third of the number of excitatory cells. Cells are placed on a 
grid (inhibitory neurons in random grid positions) of 50$\times$50 (Hodgkin-Huxley networks). Examples for lateral 
connections are indicated for an inhibitory neuron in an salt-and-pepper map (lines connecting to the neuron in the center) 
and an excitatory cell (lines connecting to the neuron at the right). The preferred orientation (PO) 
of each neuron is random as shown in the artificial orientation map. A circular Gaussian tuning curve with 
standard deviation of 27.5$^o$ is implemented. 2D Gaussian lateral connectivity based solely on 
anatomical distance between the neighboring neurons is shown on the right.}
\label{fig:mapModel}
\end{figure}

Population histogram of OSI, OMI and OI distributions for all visually responsive cells from various experimental datasets are plotted in figure \ref{fig:OSIExp}. Histograms are obtained using different measures of orientation selectivity and shows vast difference. In figure 1 (A and B) excitatory firing rate OSI and OMI admits broad range of values. Same for the firing rate OSI and OMI for inhibitory populations. Compared to figure(1A) in figure(1D)firing rate OSI distributions are much more skewed and admits 
only medium to high OSI values as reported by Tan et. al.~\cite{Tan2011}. On the contrary, firing rate population distribution from Liu data largely overlaps with figure 1(A) from Runyan.   
To check the consistency among the reported orientation selectivity data, we have estimated $p$ values using two sample Kolmogorov-Smirnov (K-S) goodness of fit test between various selectivity distributions (firing rate selectivity of excitatory and inhibitory neurons, membrane potential, excitatory conductance, inhibitory conductance of excitatory neurons). Test results with $p$ values between cross-experimental datasets are then tabulated in Table 1. From table 1 we find that the firing rate OSI of excitatory populations as reported by Tan et. al. is significantly different from Liu, Runyan, Niell et. al. ~\cite{Liu2009, Niell2008, Runyan2010} as the computed $p$ values are less than $0.001$ for all cases. Excitatory and inhibitory conductance OSI and OI distributions between Liu and Tan data respectively gives $p$ values greater than the threshold set at $0.05$. Hence, we conclude they are a good fit.
$P$ values from membrane potential OSI distributions for excitatory population between Liu and Tan data falls below 0.05 cut off and hence, the two distributions are different. $P$ values computed for firing rate OSI and OMI distributions of excitatory population between Runyan and Niell are below $0.05$ and hence, these distributions are different. On the contrary, $p$ values from inhibitory OSI distributions among the same data sets are significant. $P$ values computed for firing rate excitatory OMI, OSI distributions reported by Niell compared with Liu data are below the threshold value set at $0.05$. We conclude many of the response measures are consistent with each other, while, others are completely nonoverlapping. 

\textbf{Network with salt-pepper architecture gives spike selectivity}

\emph{Parameter space exploration reveals various dynamical regimes of salt-pepper network model}

\begin{figure}[h!]

  \includegraphics[scale=0.4]{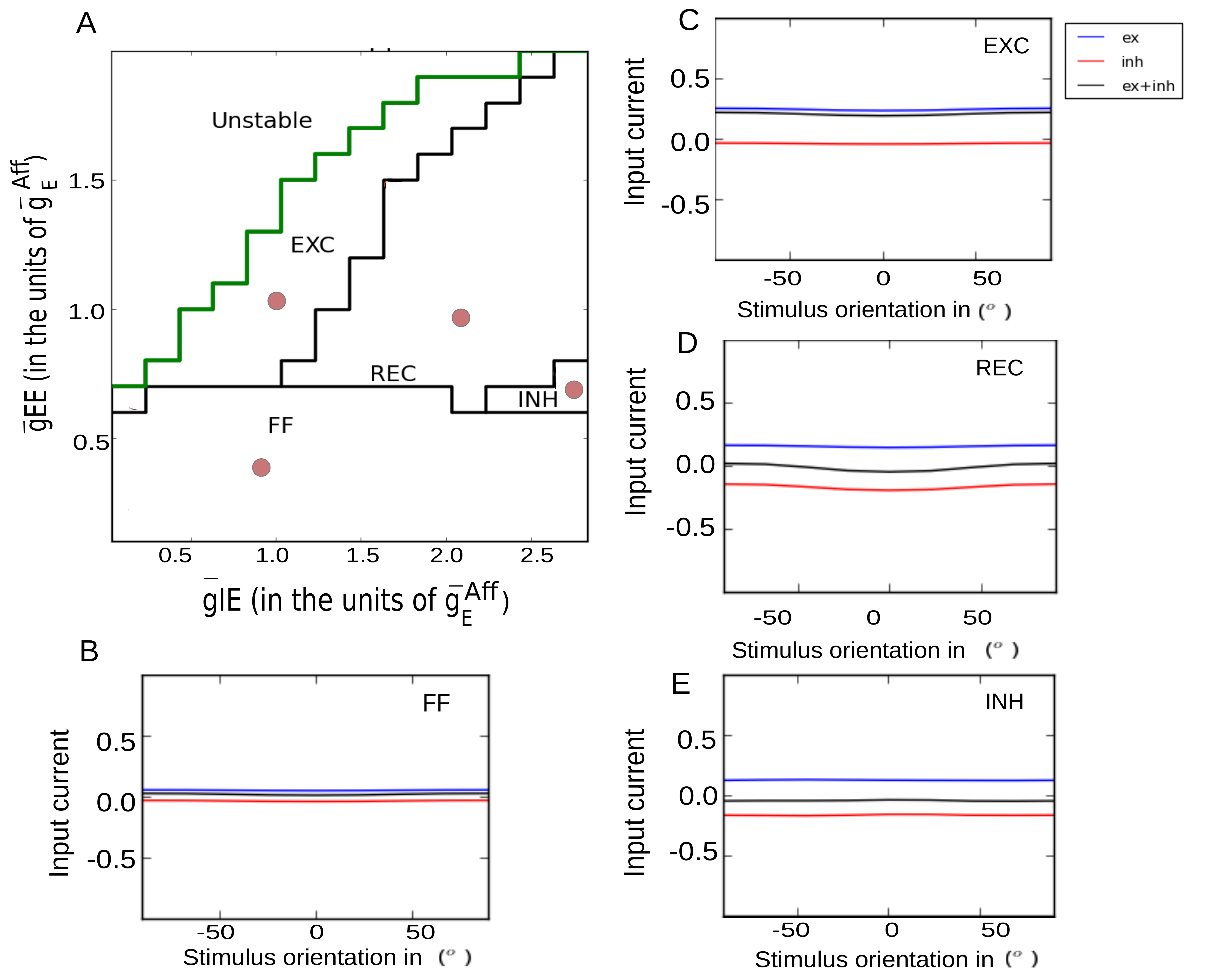}
  \caption[Operating regimes and input currents]{Operating regimes resulting from 
  19x14 different values of $\bar g_{EE}$ and $\bar g_{IE}$  in the units of maximum
  conductance of afferent input to excitatory neuron $\bar g^{Aff}_E$. 
  (a) Operating regime and regime boundaries of salt-pepper network model.
  Mean excitatory and inhibitory input currents are computed for points from each regime marked with red dot in figure \ref{fig:OperatingRegime} A and are plotted in figure (\ref{fig:OperatingRegime} B,C,D, E). 
The mean excitatory and inhibitory input currents are depicted with blue and red solid lines, respectively, normalized with the afferent input at preferred orientation. 
  The input currents at preferred orientation are aligned to 0 $^o$. The black solid lines depict the total aggregate input current.} 
  \label{fig:OperatingRegime}
\end{figure}

Network parameter space exploration results are plotted in figure \ref{fig:OperatingRegime}(A) characterizing the operating regimes and 
the various regime boundaries estimated using total excitatory and inhibitory currents received by V1 neurons, respectively. 
Each of the parameter regimes displays a characteristic relation between the OSI of the output rate membrane potential, spike, the excitatory and the inhibitory conductance, and the tuning of
the local input area (map OSI). Typical examples of these relationships are reported in \cite{Stimberg2009}.
Salt-and-pepper network shows its stability at high parameter combination of recurrent excitation, inhibition values (near the right corner) and has a small inhibitory regime. The region shown above the green solid line in figure \ref{fig:OperatingRegime}(A) corresponds to an unstable regime. In this regime the average firing rate of neurons are above 100 HZ.
\emph{Total excitatory and inhibitory input currents to V1 cells characterize parameter boundaries}

\begin{figure}[h!]

    \includegraphics[scale=0.54]{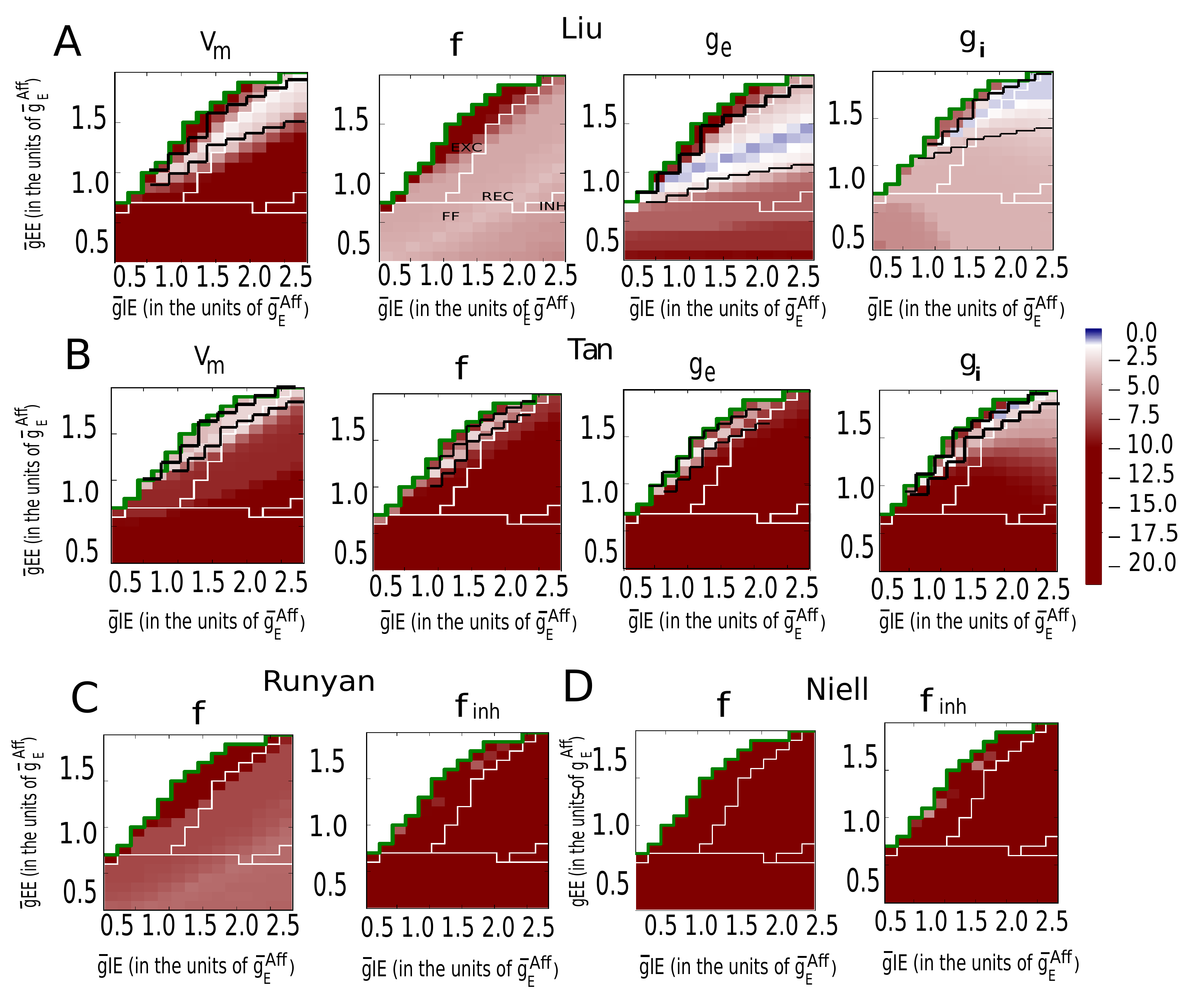}\\
     \caption[Default network p-values]{Log of p-values are plotted for spike OSI, 
      membrane potential OSI, excitatory and inhibitory OSI distributions. p-values computed using 
      two sample K-S test against experimental database. Computed p-Values are then plotted for 19x14 
      different values of recurrent excitation $\bar g_{EE}$ and $\bar g_{IE}$ in the units of maximum 
      conductance of afferent input to excitatory neuron $\bar g^{Aff}_{E}$. Null hypothesis of indistinguishable OSI distributions is validated by using a p-value threshold set at 0.05 at $95\%$ confidence interval. Colours shifted more towards light blue/white in the colormap indicates significance (p $\geq$ 0.05) and are enclosed by a boundary indiacted by black solid lines, hence, showing part of the parameter space that gives indistinguishable distributions (simulated OSI distributions vs experimental database).}     
\label{fig:Defaultparameter}  
\end{figure}

\begin{figure}
    \includegraphics[scale=0.53]{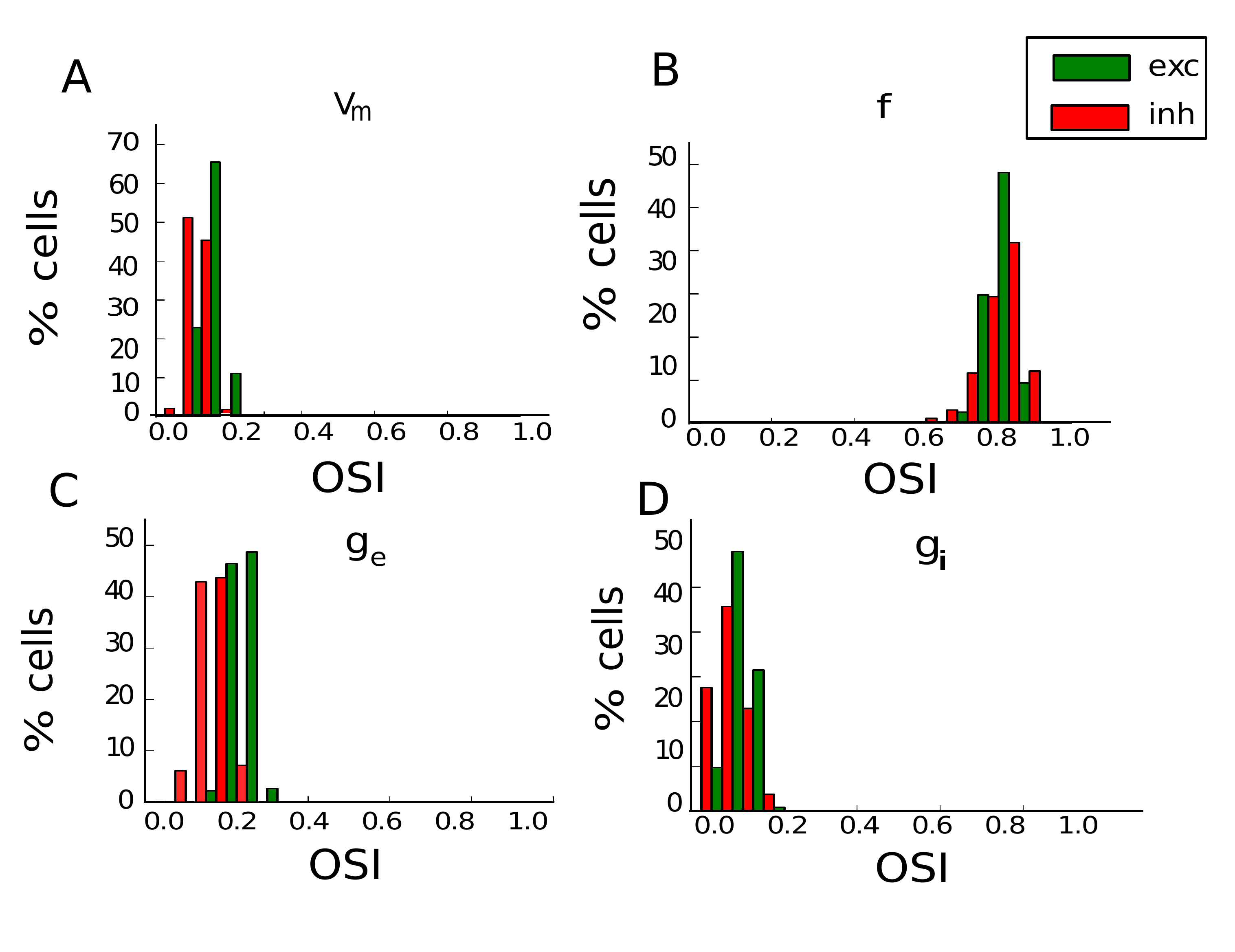}
    \caption[Tuning of excitatory neuron properties]{Histograms for inhibitory and excitatory cells are shown for a recurrent point of our network. 
$V_{m}$ OSI distribution is plotted in A, In B spike output $f$ OSI is shown and in the bottom row, C and D shows $g_{e}$, $g_{i}$ OSI distribution for the conductances.   
Both inhibitory population spike OSI (shown in red) and excitatory population spike OSI (shown in green) are highly orientation selective for the default choice of our network parameters.}  
\label{fig:TuningProp}
\end{figure}

Estimated total excitatory and inhibitory synaptic input currents (Afferent excitatory plus excitatory and inhibitory recurrent inputs measured for all V1 cells) for four arbitrary points (shown in red dot in the characterized map) of figure \ref{fig:OperatingRegime}(A)  
are then plotted in figure \ref{fig:OperatingRegime}(B),(C),(D) and (E). Blue solid line indicates total excitatory current to V1 cells, red line indicates total inhibitory currents and black line is the total aggregate current (which is a sum of total excitatory and inhibitory currents). Total input aggregate current defines the operating regime of a given network. Similar to the orientation tuning curve, the normalized input currents were aligned to the preferred orientation, which were set to 0$^o$. 
It is important to note that in the salt-and-pepper network, the cells receive collective recurrent input from their neighboring cells which have diverse orientation preferences (local map OSI). Hence, these input currents are not pronounced at the preferred orientation (0$^o$). Conceptually this is quite similar to the diverse orientation preferences of adjacent cells at any pinwheel center in a pinwheel-domain network, the recurrent input received by these cells have similarity to the cells in salt-and-pepper network.  
In our model class with random connectivity every excitatory and inhibitory cortical neuron received well-tuned time-invariant afferent input, with a reduced afferent input strength for inhibitory neurons. The synaptic strengths (maximum conductance values) of excitatory connections to excitatory ($g_{EE}$) and inhibitory cells ($g_{IE}$) were varied systematically, keeping all other parameters constant.
For each such model parameterization, we determined the best-fitting regression lines for the OSI of Vm, f, ge, and gi as a function of the local input map OSI. Subsequently, log of p-values for OSI and OI distributions $(f, V_{m}, g_{e}, g_{i})$ are validated against experimental database for each such parameterization as described above for our HH network model. 
Computed p values on a log scale are then embedded on a 2d parameter space $g_{EE}$, $g_{IE}$ as shown in figure \ref{fig:Defaultparameter} A, B and C. In the figure, the region of the parameter space bounded by black solid line provides the goodness of fit against experimental distributions. The region bounded by this black solid line assumes p values greater than 0.05 cutoff (Within $95\%$ confidence interval for a $p$ value cutoff set at 0.05). Our network as described before has an unstable regime where the average firing rate exceeds 100 Hz (above the thick solid green line), we refer to the regime as “unstable” because the network shows self-sustained activity, i. e. the network activity remains at high
firing rates if the afferent input is turned off. We do not evaluate the p values in this region.
More specifically, network simulation yield $p$ values that are above the p value threshold set at 0.05 against the Liu data sets for $(V_{m}, g_{e}, g_{i})$ OSI distributions in figure\ref{fig:Defaultparameter} A. We have a good fit for our network parameters residing mostly in a recurrent operating regime\cite{Liu_2011} for a relatively higher combination of synaptic strengths (maximum conductance values). We don't find any significance boundary against Liu's firing rate OSI distributions as indicated by figure\ref{fig:Defaultparameter} A. 
Network parameters from excitatory operating regime offers good fit against all Tan data sets.
Combination of maximum conductance values are on the higher side of the grid (see \ref{fig:Defaultparameter}B). Not too surprisingly, simulated OSI distribution for firing rates are different from experimental OSI and OMI distributions as reported by \cite{Niell2008, Runyan2010}, which is shown in the figure \ref{fig:Defaultparameter} C and D. From the above results, it is quite clear HH network model with chosen recurrent parameter grid with default network parameterization cannot capture all the relevant OSI, OMI distributions in particular for response properties such as firing rate selectivity of V1 neurons. The reported values of inhibitory spike OSI and OMI are a complete mismatch. One of the main reason for the above difference could be the following, While, Tan(2011) and Liu (2011) et. al. suggest that inhibitory cells are broadly tuned for all orientation and excitatory cells are more sharply tuned, Niell (2008), Runyan (2011) et. al.(see figure \ref{fig:OSIExp}) experimental data seems to suggest that both inhibitory and excitatory cells have a range of OSI values and hypothesize  presence of subtype of inhibitory cells in the visual cortex~\cite{Runyan2010}. Hence, the overall response specificity may be highly related to the interneuron subtype specificity and their connectivity to other cell types in the same layer of cortex.
Validation against experimental data suggest that the most likely combination of maximum conductance values ($g_{EE}$, $g_{IE}$) resides in a regime close to the boundary of recurrent and excitatory regime.    
Next, we look at the overall orientation selectivity for all $N_{E}$ = 2500 excitatory and $N_{I}$ = 833 inhibitory neurons for such a point of our network characterized by {$g_{EE}=1.5$, $g_{IE}=2.4$}.  Excitatory and Inhibitory population distributions for four response properties $(f, V_{m}, g_{e}, g_{i})$ are shown in figure \ref{fig:TuningProp}. In figure \ref{fig:TuningProp} OSI distribution of membrane potential selectivity is skewed towards zero. Intriguingly, spike output response (for excitatory neurons in green and inhibitory neurons in red) are highly selective. Both excitatory and inhibitory conductances are broadly tuned for orientation. It appears that moderate afferent tuning of both populations by and large dictates their firing rate output response. Compared to figure \ref{fig:OSIExp} we have good agreement with experimental distribution of OSI, OI values published by Liu, Tan et. al. respectively ~\cite{Liu_2011, Tan2011}. The median of the membrane potential OSI (median=0.0851) and mean of inhibition OI (mean=0.2929) are similar to their results, while the excitatory conductance OI values (mean=0.4383) is slightly higher than their result. Instead, if we select network parameter values {$g_{EE}$, $g_{IE}$} from excitatory regime of our parameter space, the median of firing rate OSI distribution resulting from simulation (median=0.3292, 0.0129 $\leq$ OSI $\leq$ 0.5588) is consistent with their result. 
The median of membrane potential OSI (median = 0.0734) is slightly lower, while the mean of excitation OI (mean=0.3626) and the mean of inhibition OI (mean=0.2400) are similar to their results. On the contrary, our simulated distribution is vastly different compared to OSI distribution of firing rates for both excitatory and inhibitory population as published by Niell, Runyan et. al.~\cite{Niell2008, Runyan2010}. Their data shows overlapping and a range of OSI, OMI values. In our data of spike output responses such diversity in orientation selectivity is not reflected inspite of the diversity present in our salt-pepper map model of V1. 
   

\textbf{Significance boundary grows with orientation dependent connectivity}

\begin{figure}[h!]
    \includegraphics[scale=0.8]{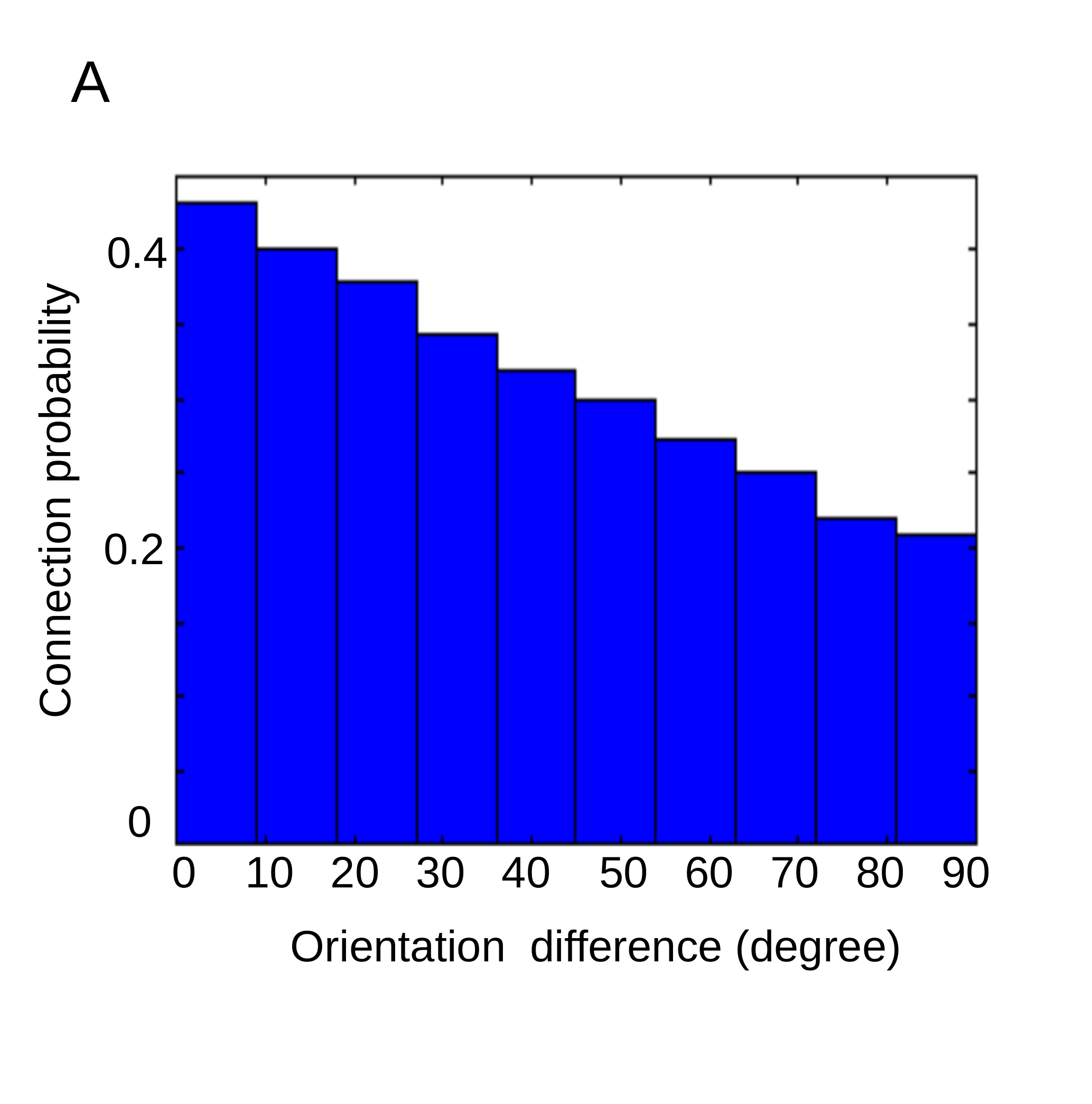}
       \caption[Fine scale connectivity results]{Connection between E-E connection pair in our network are drawn from triangular distribution. Probability of connection scales as a function of orientation difference between pre and postsynaptic cells in model V1. For example 0$^o$ orientation difference has very high connection probability (about 0.4) and 90$^o$ orientation difference has low connection probability (about 0.2).}  
\label{fig:LateralConnectivity}
\end{figure}

\begin{figure}[h!]
    \includegraphics[scale=0.46]{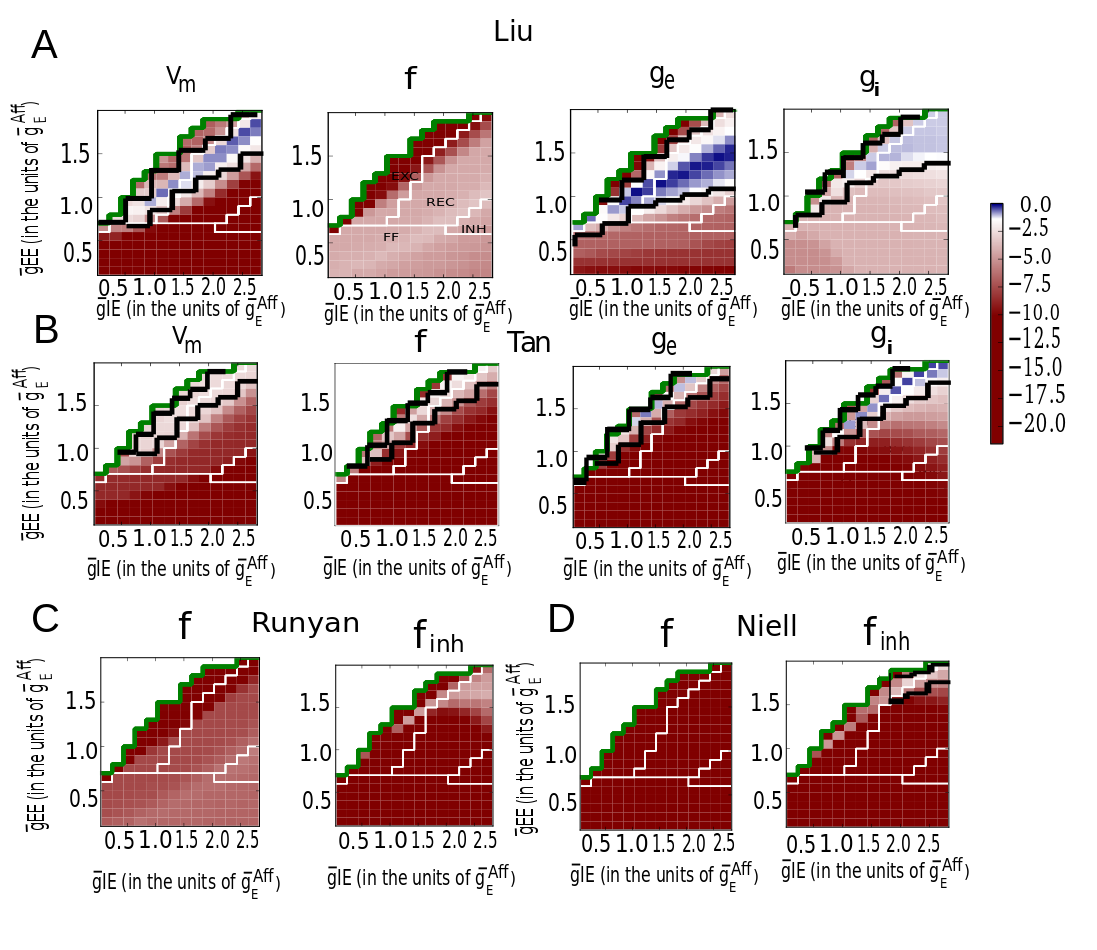}
    \caption[Full simulation of fine scale network]{Log of p-values are plotted for spike OSI, membrane potential OSI, 
    excitatory and inhibitory conductance OSI distributions with orientation dependent E-E connectivity. 
P-values are computed using a two sample K-S test against experimental database. 
Computed p-Values are then plotted on a log scale for 19x14 different combinations of recurrent excitation $\bar g_{EE}$ and recurrent inhibition $\bar g_{IE}$ in the units of maximum conductance of afferent input to excitatory neuron $\bar g^{Aff}_{E}$. In A for $V_{m}$, $g_{e}$, $g_{i}$ region enclosed within the black solid lines indicate significance (p $\geq$ 0.05). In B for $V_{m}$, $f$, $g_{e}$,$g_{i}$ parameter combinations enclosed by black solid lines show significance (p $\geq$ 0.05). In D for inhibitory spike $f_{inh}$ OSI region enclosed by black solid lines show significant values (p $\geq$ 0.05).}   
 \label{fig:FullValidation}             
 \end{figure}

\begin{figure}[h!]
    \includegraphics[scale=0.54]{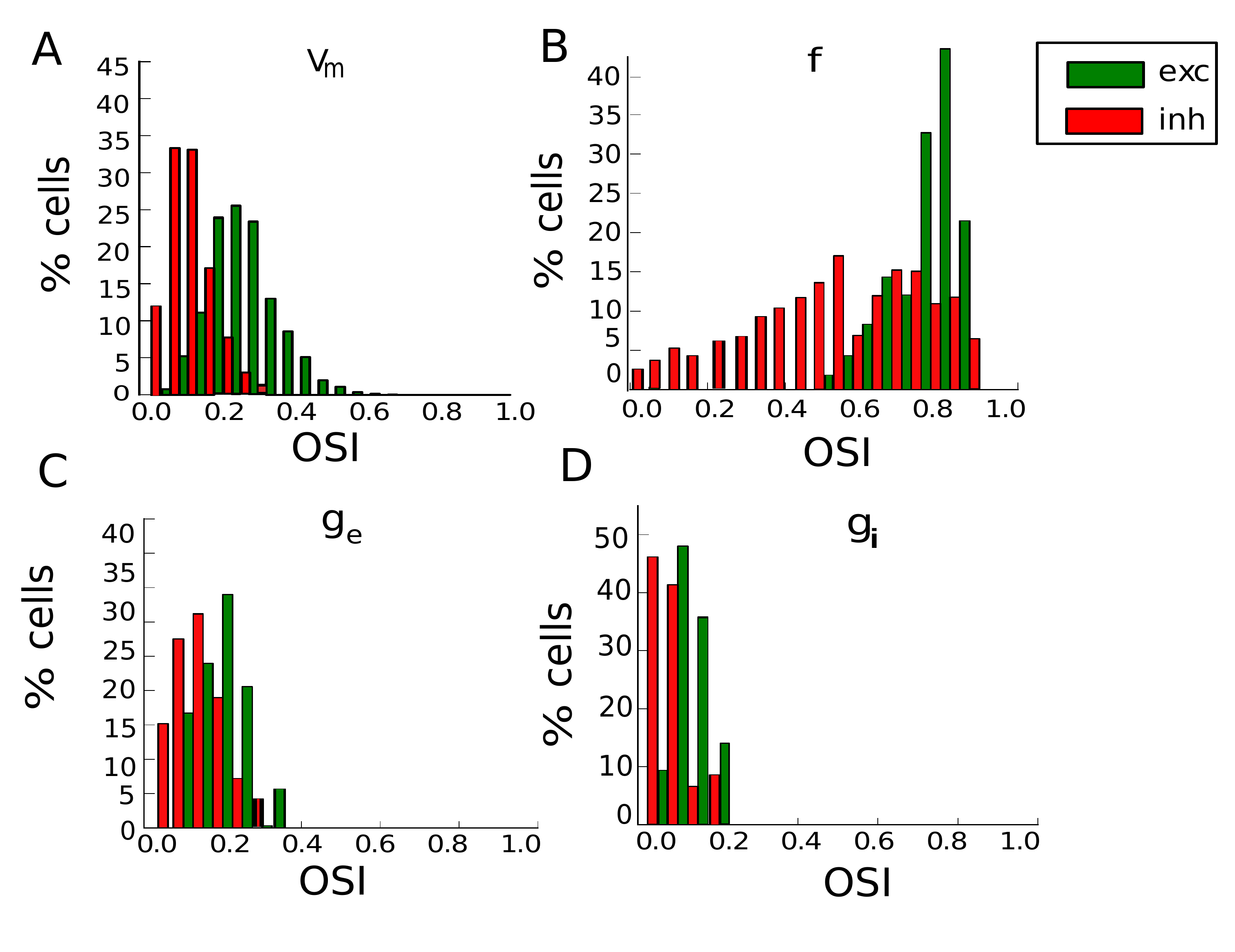}
\caption[Bar plot output OSI]{Histograms of all four response properties ($f$, $V_{m}$, $g_{e}$, $g_{i}$) are plotted using a recurrent point {$g_{IE} = 2.5$, $g_{EE}=1.5$} of the model space
with orientation dependent connectivity. Excitatory population shows high selectivity for subthreshold membrane potential and suprathreshold spike selectivity. Inhibitory OSI distributions
are broad and less selective for membrane potential and spikes.}  
\label{fig:HistogramTuning}  
\end{figure}

We argue based on the previous set of results  substantial modification in model assumption is necessary to account for maximum amount of experimental observations.
One of the plausible modification in this direction could be to incorporate a different lateral connections to extend our model space. 
To address this systematically we introduce fine scale specificity in the lateral connections based on numerous recent data published by \cite{Souza2005, Yoshimura2005, Kampa_2011}. Authors clearly suggest that in V1 most of the layer 2/3 pyramidal cells responds typically selectively to oriented grating stimuli, most probably reflecting their nonrandom connectivity. Moreover, several recent studies that combined in vivo two-photon calcium imaging with post-hoc paired
whole-cell recordings in brain slices reported evidence for functional sub-networks of neurons expressing similar orientation
tuning~\cite{Yoshimura2005, Kampa_2011}. Taking into account the above findings, We have investigated a model space with specificity in lateral connectivity between excitatory-excitatory pairs and keeping the afferent input to the cells unchanged from default parametrization. 
Local connectivity structure between exc/exc pairs in V1 is illustrated in figure \ref{fig:LateralConnectivity}(b) which has a fairly reasonable degree of resemblance with Hofers data in~\cite{Hofer_2011}. Orientation dependent connectivity as exemplified in figure \ref{fig:LateralConnectivity}(b) are drawn from a triangular distribution. In order to quantify the network selectivity of neuronal responses we compute OSI values across all excitatory and inhibitory neurons. We systematically record four different response properties membrane potential, excitatory and inhibitory conductances (subthreshold responses) and also firing rate (suprathreshold responses) OSI for all excitatory neurons respectively. 
We carry out a full network parameter search as in the previous section to validate simulated network data against experimental distributions. We use identical network parametrization as before.
Log of p-values using K-S test are plotted in figure \ref{fig:FullValidation} which extends significance boundaries in figure \ref{fig:FullValidation} A, compared to the test results for $V_{m}$, $g_{e}$, $g_{i}$, the Liu data sets plotted in figure \ref{fig:Defaultparameter} B (White and sky blue colours in our colormap corresponds to statistical significance (p $\geq$ 0.05)). Significance boundary (Within black enclosed line (p $\geq$ 0.05))also grows in 
figure(\ref{fig:FullValidation} B) in comparison with the Tan data sets in figure (\ref{fig:Defaultparameter}  B). All possible combinations of maximal conductance values $g_{EE}$, $g_{IE}$ embedded in recurrent regime of our network concurs well against Liu data $V_{m}$, $g_{e}$, $g_{i}$ OSI distribution. On the contrary, we have more excitatory points in our search grid that concurs against the Tan data for $V_{m}$, $g_{e}$, $g_{i}$, $f$. Our model space do not show any significance boundary against Liu, Runyan, Niell firing rate OSI, OMI values for excitatory population and Runyan firing rate OSI values for inhibitory population. There are few points mostly residing on the boundary of excitatory and recurrent regime that corresponds to significant values (p $\geq$ 0.05) computed for inhibitory spike OSI distributions against Niell's OMI values as shown in figure (\ref{fig:FullValidation}D). There is atleast one point from the recurrent regime of our network $g_{EE} = 1.5$, $g_{IE} = 2.54$ which can account for entire data sets of Tan and partial data set of Liu and Niell.   
We plot histograms for all response properties and for all cells in V1 for network parameter combination selected from above. Results are shown in figure (\ref{fig:HistogramTuning}). From the distribution, we find compared to figure \ref{fig:TuningProp}), in figure (\ref{fig:HistogramTuning}) membrane potential $V_{m}$ and excitatory conductance $g_{e}$ OSI distribution for excitatory cells (shown in green) have higher selectivity (mean $V_{m}$ OSI = 0.28 and $g_{e}$ OSI = 0.26). 
Same distribution for inhibitory cells  (shown in blue) show broadly tuned but a range of orientation selectivity as the distribution spreads. Mean $f$ OSI = 0.45 has lower value compared to mean $f$ OSI = 0.74 in the default case. 
Inhibitory conductance $g_{i}$ values seems to be affected the least and shows a characteristic skewed distribution for both cell types as compared to the $g_{i}$ values (mean $g_{i}$ OSI for excitatory cells $0.1368$ and mean $g_{i}$ OSI for inhibitory cells 0.0763) figure (\ref{fig:TuningProp}). Excitatory conductance and membrane potential orientation selectivity mean values are practically identical resulting in co-selectivity for stimulus orientation.  Spike selectivity for inhibitory cells are strikingly different compared to the data in 
figure (\ref{fig:TuningProp}). In this case, the inhibitory distribution is broad and has a wide range of orientation selectivity. Inhibitory population receives local recurrent input from excitatory cells (local map input OSI is diverse) in a nonorientation dependent manner and comprise of diverse tuning of excitatory conductance values. As the recurrent interactions are mediated by conductance changes therefore more spread in the excitatory conductance leads to more spread in the inhibitory spike selectivity. On the other hand, mean spike selectivity of excitatory cells are sharply tuned as they sample input from overall broadly selective inhibitory population.      
Model space with lateral connection specificity assumption definitely gives more network parameters that can provide fit against many datasets in our experimental database, however; this assumption alone is certainly not sufficient to account for may other datasets. To, investigate this further, we introduce a second modification of our network by adding an untuned component of excitatory afferent to V1 inhibitory neurons.      

\textbf{Adding an untuned component of excitatory afferent can account for maximal number of experimental observations}

\begin{figure}
\includegraphics[scale=0.52]{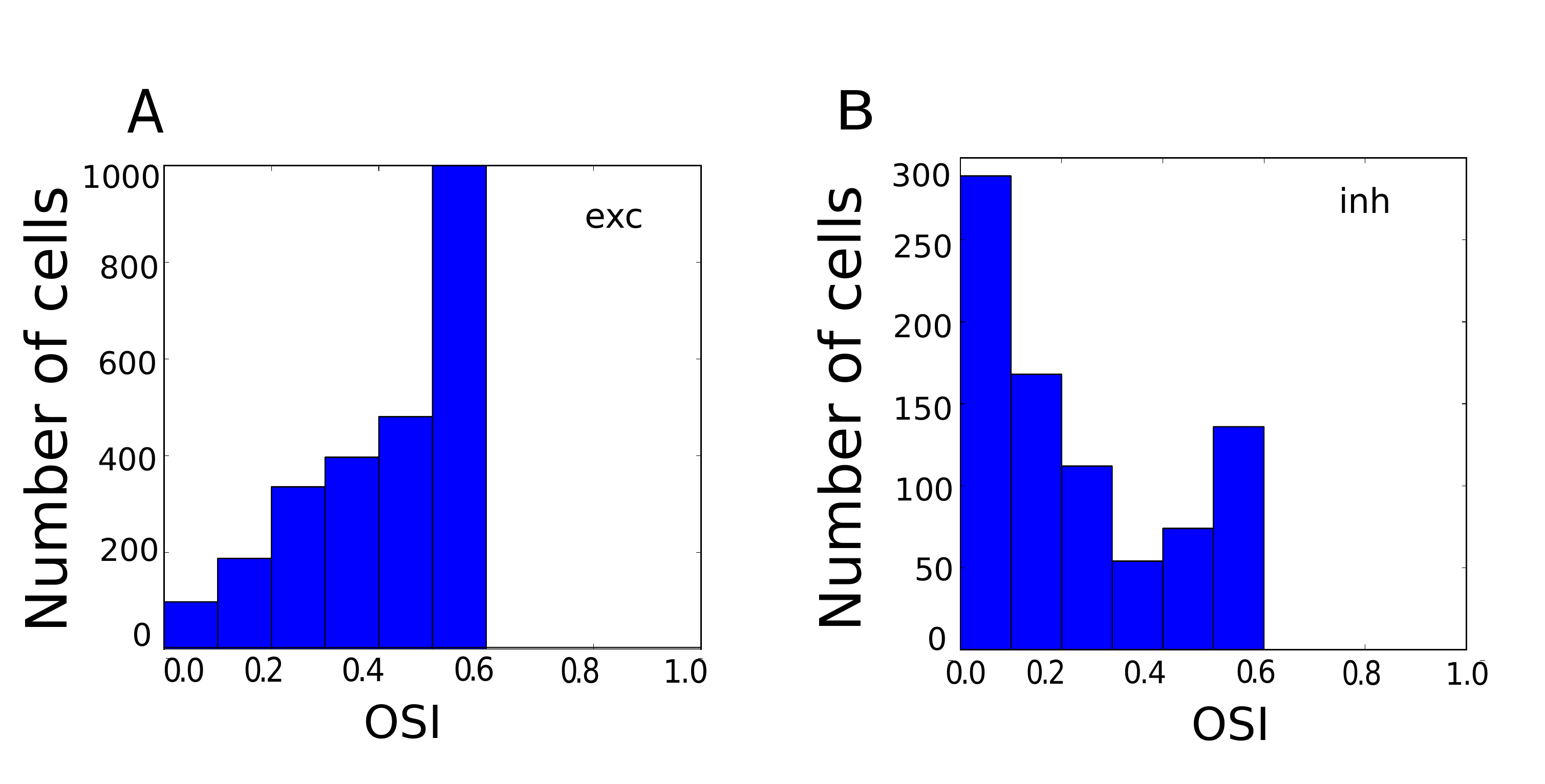}
\caption[Experimental results]{OSI distributions 
of afferent input received by excitatory cells and inhibitory
cells in V1 are plotted. 
A truncated Gaussian distribution is used with fixed mean ($\mu$) and width($\sigma$) (see text for exact values used). Inhibitory afferent distribution has 70$\%$ of neurons those are untuned as shown (cells with OSI $\ll$ 0.5).}
 \label{fig:AfferentInputBias}   
\end{figure}

\begin{figure}[h!]
    \includegraphics[scale=0.4]{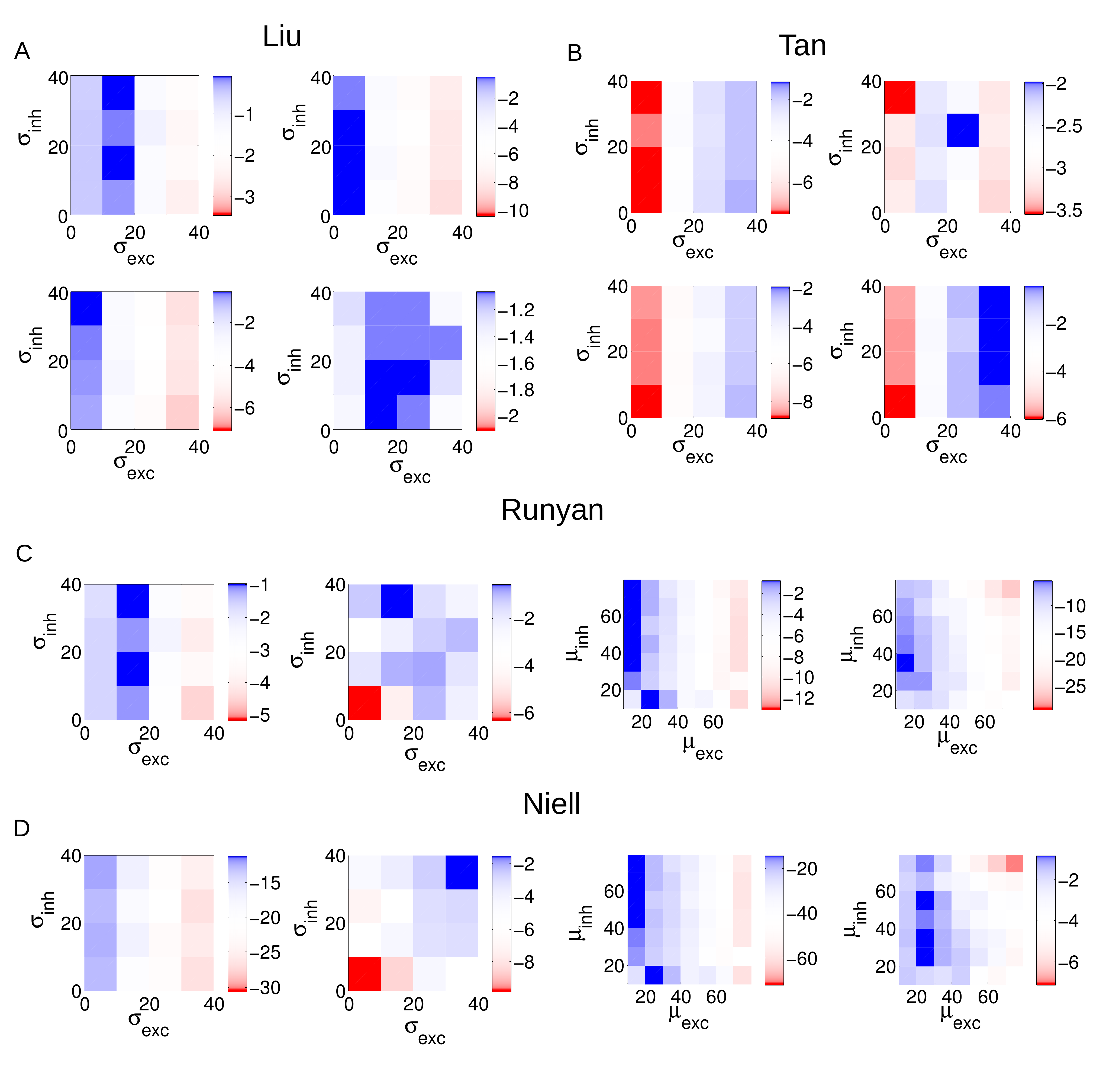}
    \caption[Experimental results]{Mean and standard deviation of a truncated Gaussian is varied. Optimal values for $\mu$, $\sigma$ is then decided based on K-S test to draw excitatory and inhibitory afferent input distributions. For figure (\ref{fig:TuningProp1} A, B) $\sigma_{exc},\sigma_{inh}$ is varied keeping mean $\mu$ of the truncated Gaussian distribution fixed. 
For figure (\ref{fig:TuningProp1} C, D), both $\sigma_{exc},\sigma_{inh}$ and $\mu_{exc}$, $\mu_{inh}$ is varied by holding other two parameters. In A Log of p values are computed against OSI distributions reported by Liu et.al. for spike, membrane potential, excitatory and inhibitory conductance selectivity for all excitatory cells. In B Log of p values are plotted using data published by Tan et. al. for spike, membrane potential, excitatory and inhibitory conductance selectivity for all excitatory cells. In C log of p values are plotted using data published by Runyan et. al. for spike selectivity of excitatory and inhibitory cells. In D Log of p values are plotted using data published by Niell et. al. for spike selectivity for excitatory and inhibitory cells. Blue/white colours in our chosen colormap indicates significance (p $\geq$ 0.05)}  
\label{fig:TuningProp1}
\end{figure}

Model classes described in the previous two sections assume feedforward afferent input to all V1 cells are strongly co-tuned for orientation. However, recent studies based on the response properties of layer 4 (a major thalamocortical input layer of V1) interneurons suggests functional differences between two inhibitory subtypes somatostatin (SOM) and paravalbumin (PV) neurons\cite{Ma_2010}. Their results suggest that SOM and PV have differential tuning in layer 4: while PV neurons are untuned for orientation, SOM neurons are as sharply tuned as excitatory neurons\cite{Ma_2010, Mao_2011}.
Here, we have hypothesized that all the excitatory cells and a fraction of the inhibitory cells in V1 receive co-tuned excitatory afferent input and the remaining fraction of the inhibitory neurons receive weakly tuned afferent excitatory input. 
For our proposed feedforward architecture, we have used $N_{u} = 70\%$ weakly tuned and $N_{t}=30\%$ tuned excitatory afferent input in order to introduce differential tuning to inhibitory neurons in V1 (see materials and methods).
To make it more transparent, orientation selectivity of afferent input received by excitatory and inhibitory neurons are shown in figure \ref{fig:AfferentInputBias}. Due to the presence of tuned and untuned components, afferent input to V1 interneurons has a multimodal distribution as shown in figure (\ref{fig:AfferentInputBias} B). 
Afferent mean firing rates are computed using a probability of distribution based on a truncated Gaussian of fixed mean and width as defined in Eq.(11). In order to reduce free parameters and find optimal values for $\mu$ and $\sigma$ we carry out a coarse grid search. We vary
$\mu_{exc}$ and $\mu_{inh}$ from 0 to 80 in a step size of 20 and held $\sigma_{exc} = 27.5$ and $\sigma_{inh}=27.5$ fixed. Subsequently, we compute the firing rate of all cortical excitatory, inhibitory cells and compute corresponding p values based on a K-S test against experimental datasets for OSI, OMI distributions from Runyan, Niell. Log of p values are then plotted on a 2D grid as shown in fig.\ref{fig:TuningProp1} D. Based on the grid search p values are most significant in a range of $\mu_{exc} = 10-20$ and $\mu_{inh} = 35-40$. Similarly, we carry out a grid search over $\sigma_{Ex}$ and 
$\sigma_{inh}$ by holding other two $\mu$ parameters fixed and also, resorting to default network parametrization. Results of log of p values computed against Liu, Tan and Runyan values are shown on this grid in figure.(\ref{fig:TuningProp1} A, B, C). Liu data in figure (\ref{fig:TuningProp1} A) indicates that optimal values for $\sigma_{Ex}$ between 10-30 and that for $\sigma_{inh}$ varies over a range of values. On the contrary, against Tan data sets we get optimal values for higher combination of $\sigma_{exc}$ and $\sigma_{inh}$ respectively. We set optimal $\sigma_{exc} = 22.5$, $\sigma_{inh}=40$ based on the rationale that this is one of the combination that provides goodness of fit against (p $\geq$ 0.05)(blue/white colours corresponds to higher significance in the colormap used here) maximum number of experimental observations. Afferent input OSI distributions are drawn with the following four input parameters; $\sigma_{exc} = 22.5$, $\sigma_{inh}=40$, $\mu_{exc}=20$, $\mu_{inh}=35$ respectively.    
In order to quantify selectivity of our V1 network with this model space we compute OSI values as before for spikes (for both excitatory and inhibitory neurons), membrane potential, excitatory and inhibitory conductances (subthreshold selectivity) for all excitatory neurons. Subsequently, we repeat our analysis by performing a two sample K-S test against experimental OSI distributions for five different response properties. We perform this test for all parametrization of recurrent excitation and inhibition values $g_{EE}, g_{IE}$.   

\begin{figure}[h!]

    \includegraphics[scale=0.45]{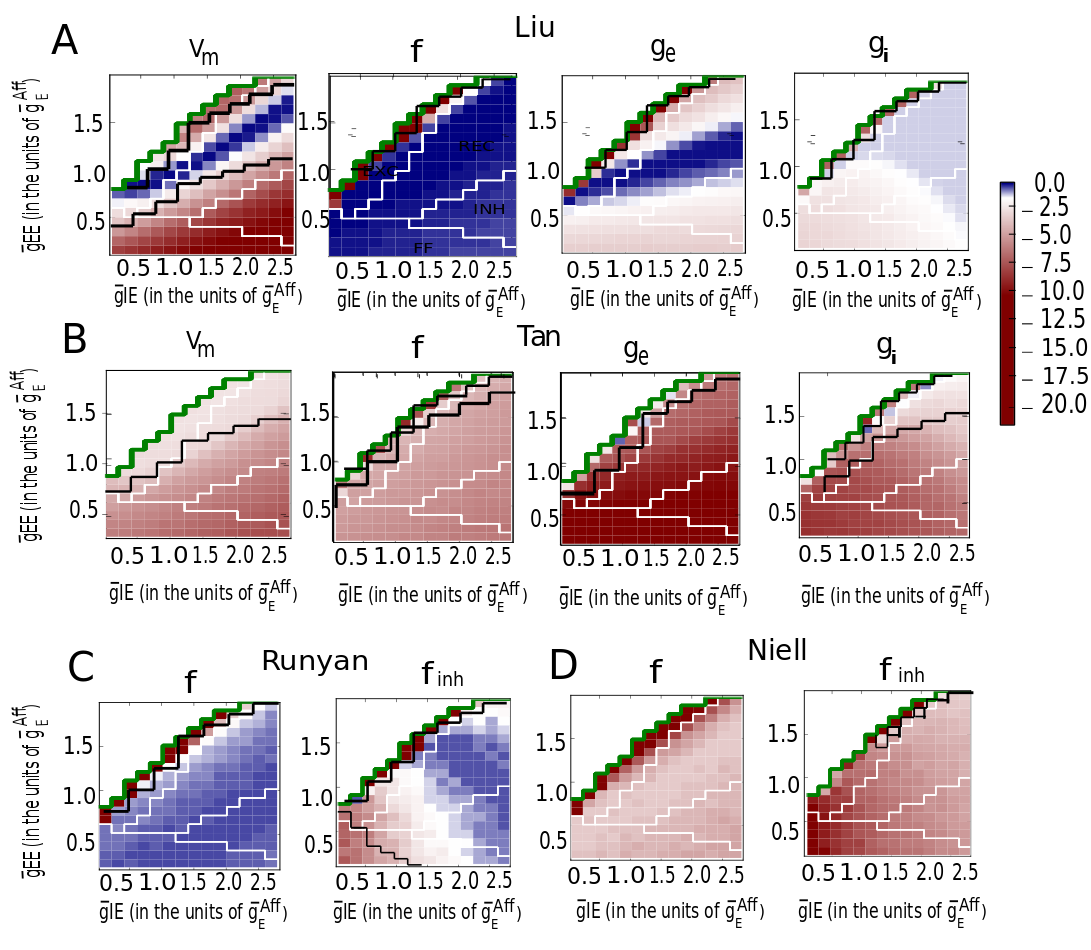}\\
     
    \caption[KS test of OSI distributions]{ Log of p-values are plotted for spike OSI, membrane potential OSI, 
    excitatory and inhibitory conductance OSI distributions with afferent specificity. 
    P-values are computed using a two sample K-S test against experimental database. 
    Computed p-Values are then plotted on a log scale for 19x14 different values of recurrent excitation 
  $\bar g_{EE}$ and recurrent inhibition $\bar g_{IE}$ in the units of maximum conductance of afferent input to excitatory neuron $g_{E}^Aff$. In A for $V_{m}$ region enclosed within the black solid lines indicate significance (p $\geq$ 0.05). For, $f$, $g_{e}$, $g_{i}$ all parameter combinations below the black solid line is significant (p $\geq$ 0.05). In B for $V_{m}$, $g_{e}$ parameter combinations above the black solid line shows significance (p $\geq$ 0.05). For $f$, $g_{i}$ significant parameter combinations are again enclosed by black solid lines. In C for excitatory $f$ region under black solid lines shows significant values (p $\geq$ 0.05) and for inhibitory $f$ most significant part of the parameter space is enclosed by boundaries. In D for inhibitory $f$ only few points above the marked boundary satisfies p $\geq$ 0.05} 

\label{fig:KS-test}
\end{figure}
\begin{figure}[h!]
    \includegraphics[scale=0.51]{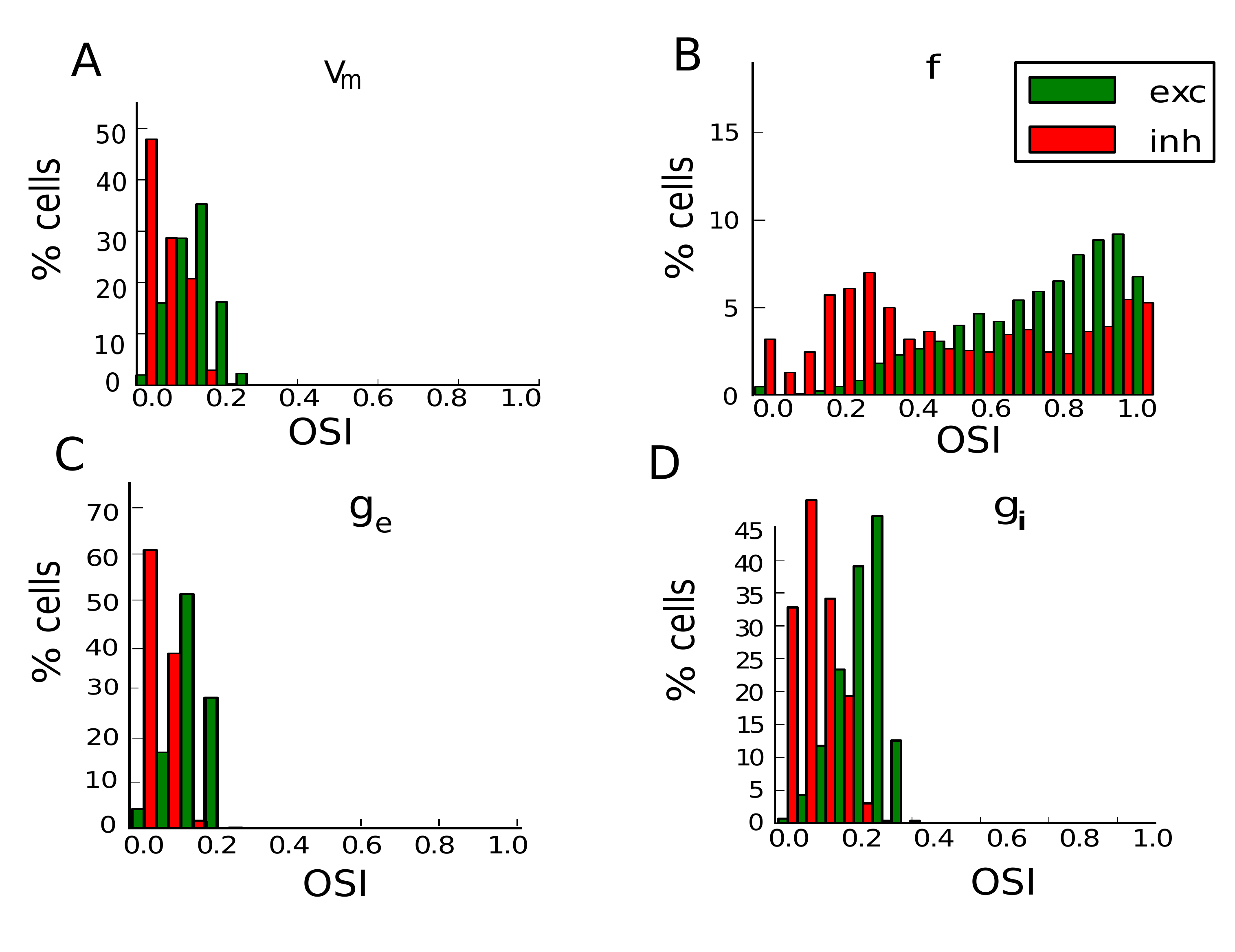}
    \caption[Experimental results]{Population histogram data for $f OSI$, $v_{m} OSI$,$ge OSI$, $gi OSI$ 
of excitatory and inhibitory cells for a recurrent point (within significance boundary) are shown in A, B, C, D respectively. Distribution use default number of $N_{EE} = 100$, $N_{IE} = 50$ connections in a recurrent network with afferent specificity. A multimodality is observed in the inhibitory distribution of OSI (in red) in B}  
\label{fig:TuningProp2}
\end{figure}

As in the previous section those log of p values which are (p $\geq$ 0.05 ) significant are enclosed with black solid line. 
In figure (\ref{fig:KS-test} A) log of p values on the 2D grid of network parameters $g_{EE}$, $g_{IE}$ can account for Liu $V_{m}$, $f$, $g_{e}$ and $g_{i}$ OSI distributions mostly at the recurrent operating regime of our network. There are many points spanning different operating regime that provides goodness of fit against Liu firing rate distributions. In fact, almost all the points below black solid line indicates significance (p $\geq$ 0.05). There is atleast one point $g_{EE} = 1.45, g_{IE}=2.45$ which satisfies the entire data set and close to the point $g_{EE} = 1.5, g_{IE}=2.5$ we found previously in our investigation with lateral connection specificity.  
Compared with figure \ref{fig:Defaultparameter} B and \ref{fig:FullValidation} B this class of model has larger area of network parameters that can account for $V_{m}$ OSI (All significance points above black solid line in figure (\ref{fig:KS-test}B)).  
Figure (\ref{fig:KS-test} B) Tan data shows network combinations are mostly from excitatory regime for $f$, $g_{e}$ and $g_{i}$ which is in line with the findings from previous two model space. Most significant areas of the parameter space is color coded in blue/white. Excitatory and inhibitory spike OSI distribution from Runyan in figure (\ref{fig:KS-test} C) is significant (p $\geq$ 0.05) in the entire area below the black solid line, and in the enclosed region; in the inhibitory plot. Different operating regimes of our network namely (excitatory, recurrent and inhibitory) commensurates with experimental distributions from Runyan et. al.(\cite{Runyan2010}).  
Network parameters in the high excitatory regime concurs against Niell's data for inhibitory $f$ OMI values. We don't find in the area searched almost no points that concurs well with excitatory $f$ OMI data. The point which provides goodness of fit across studies is situated almost at the boundary of recurrent, excitatory and unstable regime; it's maximal conductance value corresponds to $g_{EE} = 1.56$, $g_{IE}= 2.5$.  
One of our key finding is shown in figure \ref{fig:TuningProp2}, that the presence of an untuned component of excitatory afferent predicts a range of orientation selectivity for both excitatory and inhibitory populations in layer 2/3 of primary visual cortex. Results of the output population histograms for different sub and suprathreshold response properties are shown. Inhibitory spike selectivity show multimodal distributions suggesting existence of both broad and highly selective responses among these cell types (mean $f$ OSI = 0.72 for excitatory cells and mean $f$ OSI = 0.50 for inhibitory cells ). Based on the afferent input distribution shown in figure (\ref{fig:AfferentInputBias} A, B) it further enhances the dependence of V1 output responses via afferent specificity alone. Output responses correlates with specificity of afferent input.
Excitatory neurons spike OSI also have a range of values suggesting they pool recurrent inputs from inhibitory neurons of diverse orientation selectivity. Distribution shown in figure \ref{fig:TuningProp2} reproduce experimental distribution of spike selectivity (inhibitory and excitatory spike OSI in \cite{Runyan2010}) sufficiently well as can be seen from visual inspection. Mean excitatory OSI = 0.72, mean inhibitory OSI = 0.50 values also overlaps well with their reported values. Compared to spike selectivity data subthreshold response properties such as membrane potential ($V_{m}$ OSI), excitatory ($g_{e}$ OSI), inhibitory ($g_{i} OSI$) are all broadly selective and assumes values that are comparable with the mean values that resulted from the simulation of previously described model assumptions.
\textbf{Afferent specificity with recurrent fine scale connectivity can account for experimental observations} 

\begin{figure}[h!]
    
    \includegraphics[scale=0.56]{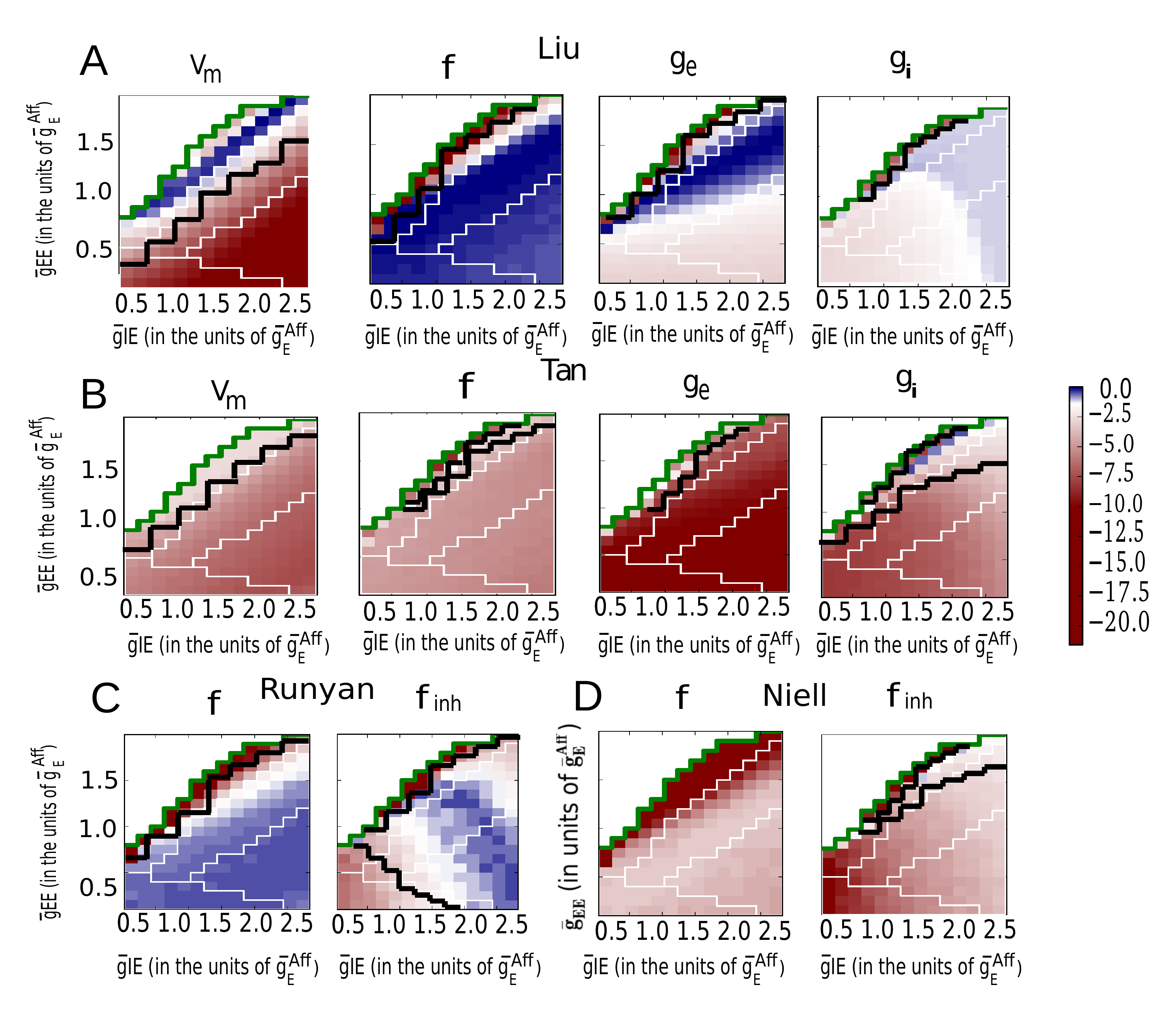}\\

    \caption[KS test of OSI distributions]{ Log of p-values are plotted for spike OSI, membrane potential OSI, 
    excitatory and inhibitory conductance OSI distributions with afferent specificity and fine scale
    recurrent connectivity. 
    P-values are computed using a two sample K-S test against experimental database. 
    Computed log of p-Values are then plotted for 19x14 different values of recurrent excitation 
    $\bar g_{EE}$ and $\bar g_{IE}$ in the units of maximum conductance of afferent input to 
    excitatory neuron $g^{Aff}_{E}$. In A parameter combinations
    above the black solid boundary provides significance (p $\geq$ 0.05) against Liu $V_{m}$ OSI distributions. For Liu $f$, $g_{e}$, $g_{i}$ all network parameter combinations below the black solid line shows significance (p $\geq$ 0.05)(shown in blue/white colours). 
In B for Tan $V_{m}$, $g_{e}$ data grid points above the black solid line provides significance. For $f$, $g_{i}$ points within the enclosed black solid lines provides significance (p $\geq$ 0.05). In C all the points below the black solid line shows significance (p $\geq$ 0.05) against excitatory spike selectivity data from Runyan et. al. and all the points within the enclosed black solid lines provides significance (p $\geq$ 0.05) against inhibitory spike selectivity. In D all the points within the black solid lines indicate significance (p $\geq$ 0.05) against Niell inhibitory spike selectivity} 
\label{fig:Networkstate}    
\end{figure}

\begin{figure}[h!]
    \includegraphics[scale=0.56]{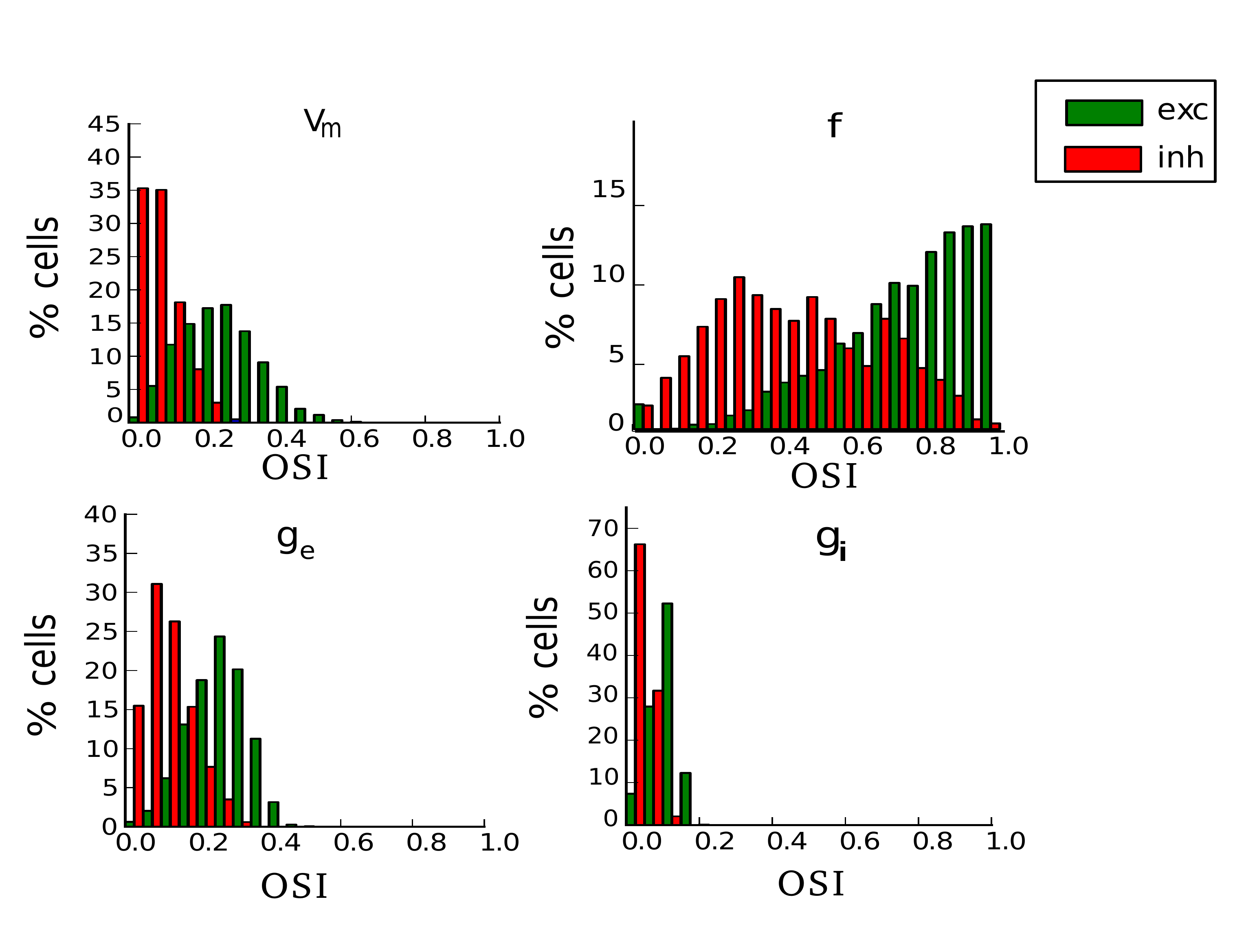}
    \caption[Experimental results]{Population histograms for $f$ OSI, $v_{m}$OSI,$ge$ OSI, $gi$ OSI 
    of excitatory and inhibitory population for a recurrent point of our model space with afferent specificity and feature specific connectivity are plotted in A, B, C, D. Distribution use default number of $N_{EE}$, $N_{IE}$ connections. In green, excitatory population distribution and in red, inhibitory population distribution are shown}  
\label{fig:FullNetworkAfferentHistogram}
\end{figure}

Afferent specificity assumption is investigated with default network parametrization. Now, we would like to extend this by incorporating two assumptions and extending our network model space.  
Two assumptions are afferent specificity in combination with lateral connection specificity. 
To this end, We first compute the p values as before between simulated and experimental OSI distributions for five different response properties ($f$ excitatory, $f$ inhibitory, $V_{m}$, $g_{e}$, $g_{i}$). Subsequently, computed log of p values are plotted on a 2D grid of network parameters {$g_{EE}$, $g_{IE}$} as shown in figure (\ref{fig:Networkstate} A, B, C, D).   
Compared with previous network validation shown in figure (\ref{fig:Defaultparameter}A) and (\ref{fig:FullValidation} A) model space for values taken from membrane potential $f$, $V_{m}$, $g_{e}$ and $g_{i}$ response 
data from Liu concurs well. All the areas under the black solid line in the figure for $f$, $g_{e}$, $g_{i}$ gives significant values (p $\geq$ 0.05) which spans almost three regions in the parameter space. Compared to figure \ref{fig:KS-test} A, spike selectivity data shows higher log of p values indicating network with afferent specificity has a better match against Liu's spike selectivity distribution. However, this network class gives slightly improved match against $g_{e}$, $g_{i}$ OSI distribution data from Liu as can be seen from the log of p values in figure \ref{fig:Networkstate} A in the recurrent
regime of our network. For, $V_{m}$ OSI data this model space shifts the significant values more towards excitatory regime. We also find this interesting shift in $g_{e}$, $g_{i}$ data. For $g_{i}$ data, parameter region with significant p values shrinks. Network seems to give more significant p values for higher combination of {$g_{EE}$, $g_{IE}$} in the recurrent regime of the network. Compared to figure (\ref{fig:Defaultparameter} B) and (\ref{fig:FullValidation}B) this class of model gives improved p values against all four response properties by Tan et. al.\cite{Tan2011}. However, compared with figure (\ref{fig:KS-test} B) there is no significant qualitative improvement in the figure (\ref{fig:Networkstate}B). Figure (\ref{fig:Networkstate}C and D)  shows spike selectivity for the entire network. Spike selectivity for excitatory cells from Runyan et. al. \cite{Runyan2010} is significant (p $\geq$ 0.05) over various dynamical regime. On the other hand, inhibitory spike selectivity is significant mostly in the recurrent and inhibitory regime of this network (far less region of significance compared to the excitatory spike selectivity). Niell's data for spike selectivity of inhibitory cells shown in Figure (\ref{fig:Networkstate} D) show more significant $p$ values for high parameter combinations mostly extracted from excitatory and recurrent regime of our network. 
This model space shows small improvements in the log of p values for spike selectivity of inhibitory cells against Niell's data, however, slightly deteriorated log of p values compared to spike selectivity data by\cite{Runyan2010}. 
Population histograms in figure (\ref{fig:FullNetworkAfferentHistogram} A, B, C, D) are plotted for one of the recurrent point {$g_{EE} = 1.5$, $g_{IE} = 2.5$} of our network which satisfies maximal constraints (orientation selectivity distribution for intercellular responses from various data sets). Histogram of spikes for excitatory and inhibitory population show substantial selectivity. Distribution has lesser proportion of inhibitory neurons which are very highly
selective as compared to the spike distribution data in figure \ref{fig:TuningProp2}. As a result, mean inhibitory spike OSI $f_{OSI} = 0.4361$ is much lower compared to the mean value shown in figure \ref{fig:TuningProp2}. This could be due to increase in recurrent input from neighboring cells with diverse orientations. There is also a slight increase in the mean spike OSI of excitatory population and likely due to higher proportion of excitatory neurons with very high selectivity. Fine scale recurrent connectivity increases excitatory conductance,  mean $g_{e}$ OSI value for excitatory neurons as shown in figure \ref{fig:FullNetworkAfferentHistogram} compared with figure \ref{fig:TuningProp2} (with green bar). As E to E interactions are mediated via synaptic conductance changes, high selectivity in $g_{e}$ OSI also show enhanced spike selectivity for the excitatory population.

\section{Discussion}
\emph{Exploration and validation of model parameter space, regime with heterogeneity in the datasets}

In this article, we assessed, which cortical operating regime and model space based on biologically plausible assumptions are more well-suited to explain the dependence of orientation tuning and selectivity properties of mouse primary visual cortex. Many recent experiments Runyan et. al. \cite{Runyan2010}, Tan et. al. \cite{Tan2011} with intracellular measurements quantify for the same neurons: (1) input (total excitatory and the total inhibitory conductance), (2) state (subthreshold tuning of the membrane potential), and (3) output (spike response). We selected this particular data set because access to both, sub-and-superthreshold tuning, recorded for the same neurons with high laminar specificity. A key to constraining our network model space. Spike output response (for excitatory and inhibitory population) alone, for example, constrained several model space and network parameter values within a particular model space. Moreover, as shown in figure (\ref{fig:Networkstate} C) spike output is compatible with wide range of excitation and inhibition values. This is also true for investigation with afferent specificity assumption alone. We systematically varied the strength of recurrent excitation and inhibition values relative to the strength of the afferent 
conductance values in a biologically more plausible Hodgkin-Huxley network model. We have investigated network parameter combination of recurrent excitation
 and inhibition values embedded in a specific region of the parameter space those satisfies maximum number of data constraints based primarily on orientation selectivity.
For all types of models, our finding suggests that recurrent regimes dominated by relatively high maximal conductance values can support maximum number of experimental distributions. These are important to know for reproducing the experimental results. 
This regime consist of excitatory and inhibitory components balancing one another and dominating the afferent input drive. Due to the strong recurrent drive, the most likely excitation and inhibition values are located close to a region in parameter space where the network settles into a state of strong, self-sustained activity.
Recently Hansel et.al. have shown that V1 cortical networks without having any reliance on a functional map can generate sharp selectivity while operating on a balance regime of the network, 
even though anatomically close cells have diverse POs~\cite{Hansel2012}. 
In our network investigations with a salt-pepper map model, in which the lateral connectivity depends only on anatomical distance, the excitatory and inhibitory population show high selectivity close to the recurrent operating regime for Liu datasets (see for $(V_{m}, g_{e}, g_{i})$ OSI distributions in figure (\ref{fig:Defaultparameter} A) ) but for the Tan datasets 
figure (\ref{fig:Defaultparameter} B), network satisfies maximal number of constraints in an excitatory regime. 
Moreover, afferent input specificity to V1 inhibitory neurons allowed us to test directly to what degree emergence of orientation selectivity is largely independent to the degree of orientation selectivity in the input layer as found by Hansel and van Vreeswijk (2012) et. al.\cite{Hansel2012}.
We find on the contrary, spike output responses are highly dependent on the input tuning. Our null hypothesis about the afferent specificity is crucial. In fact, adding an untuned component to excitatory afferent input to V1 inhibitory population can account for majority of the experimental data discussed here.  
Clearly, the spike output responses for both excitatory and inhibitory population as shown in figure (\ref{fig:TuningProp2}) correlates with the structure and type of distribution of their respective inputs (unimodal for excitatory and bimodal for inhibitory) (see figure (\ref{fig:AfferentInputBias}A and B)). Recurrent connectivity plays a less dominant role in this model space. Input tuning by and large dictates output responses. This is further raising the possibility of an alternative mechanism for orientation selectivity of V1 neurons in mouse visual cortex.

\emph{Afferent specificity to inhibitory population in v1 modulates visual response}

Inhibition play critical functions in shaping spontaneous and
evoked cortical activity, and hence inhibitory interneurons
have a key role in emergent properties such as orientation 
selectivity in cortical circuits ~\cite{Liu2009, Runyan2010, Niell2008}. 
In the visual (and other sensory) cortex, inhibition appears to have a 
very important role in influencing the receptive field properties of neurons.
The role of inhibition in shaping response features such as
orientation selectivity \cite{Hirsch_2003, Lauritzen_Miller_2003} 
has been difficult to resolve unequivocally, due in part to multiple
nonlinear properties (importantly, the spike threshold) available 
to cortical neurons for integrating inputs and generating
selective responses~\cite{Tan2011}. 
The tuning of inhibitory inputs to an individual neuron is
variable \cite{Kuhlmann_Vidyasagar_2011, Monier_Chavane2003} 
and depends at least partly on the location of neurons within 
relevant feature maps \cite{Marino2005}.
Even less is known about the role of inhibition in the mouse
visual cortex, which lacks orientation columns~\cite{Dräger_1975, VanHooser2006} and may therefore require additional inhibitory mechanisms. 
Orientation tuning of inhibitory neurons in mouse visual
cortex has been described as generally broader than that of
excitatory neurons \cite{Liu_2011, Liu2009, KerlinReid_2010}. 
However, careful analysis of paravalbumin-expressing interneurons 
using targeted recordings from single marked cells demonstrates that 
they have a range of tuning properties, including very sharp orientation selectivity~\cite{Runyan2010, Niell2008}. We have proposed here a model hypothesis with afferent specificity, which somewhat related to what is suggested by recent physiological studies, which has emphasized the role of both a tuned and untuned inhibitory component in the generation of orientation selectivity \cite{Lauritzen_Miller_2003, Shapley}, that could be realized in layer 4 by two functionally different groups of inhibitory interneurons \cite{Hirsch_2003, Nowak_2008}. 
We have tested the influence of an untuned afferent input component on orientation selectivity and tuning of excitatory neurons in V1.
We find by marginalizing inhibition in this way a range of orientation
tuning for V1 inhibitory neurons can be fitted which is consistent 
with the findings from paravalbumin-expressing interneurons by Runyan et. al. \cite{Runyan2010}. 
Further, our computational model guarantees in a sparsely connected network presence of differential input tuning to inhibitory cortical cells is sufficient for explaining diverse orientation selectivity of cortical cell types. In general, dendrite targeting vs soma targeting subclass of interneurons may have a precise functional role in dividing orientation selectivity of neurons in V1. It would be pertinent to investigate this in a more complex class of model having multi compartments to properly characterize the functional role of interneuron cell types. This is 
indeed one of the shortcoming of our model fitting approach.
 However, we can still show specific modulation of visual responses of 
V1 cell types due to the presence of differential tuning of cell types (excitatory and inhibitory)in the same local circuit.
  
\emph{Contribution of broadly tuned inhibitory neurons on contrast invariant tuning}

Early modeling studies proposed mechanisms such that inhibition from neurons with orthogonal orientation preferences (Wörgötter and Koch, 1991; Sabatini, 1996) might suppress activation at nonpreffered orientation\cite{Wörgötter_Koch_1991, Sabatini_1996}. Similar proposed mechanisms relied on broadly tuned (Troyer et al., 1998) or untuned (Lauritzen and Miller, 2003) inhibition\cite{Troyer_1998, Lauritzen_Miller_2003}.
Although, we haven't shown here any data for contrast invariance of tuning width; we report here briefly that our preliminary findings in the mouse visual cortex model space with afferent specificity
is consistent with this early proposals.
In general, in the present implementation of our model afferent input tuning width is independent of contrast. However, effect of stimulus contrast can be incorporated in our network readily by altering the baseline of afferent firing rate of $30$ Hz by systematically changing the contrast dependent factor in Eq.(9). 
The increase of stimulus contrast in the network model with afferent specificity increases the afferent input to the neurons in V1, which might increase the spike response at non-preferred orientation. The broad or almost untuned inhibition resulting from widening of the tuning width of afferent input to inhibitory cells in V1 is capable to suppress this spike response at non-preferred orientation. 
The networks with broader standard deviation of afferent input $\sigma_{inh}$ are more contrast invariant compared to the others. Thus, this observation would agree with the hypothesis by \cite{Liu2011c}, which stated that the contrast invariant in mouse V1 is achieved primarily with contrast-dependent modulation of inhibitory tuning strength. We find in our data (not shown) most of the tuning curves exhibit their contrast invariant at least by 50\% contrast.
Hodgkin-Huxley salt-and-pepper model space with default parametrization also entails the possibility for contrast invariant tuning. 
More recently, it has also been proposed that not the network connectivity but the properties of the transfer function, converting synaptic inputs into firing rate are responsible for achieving contrast invariant tuning. In vivo transfer functions are well
approximated by power-law functions (Anderson et al.; Finn et al.) \cite{Anderson_2000, Finn_2007} and similarly, transfer functions in the Hodgkin-Huxley model with default parametrization of the network presented in this study are also well described by a power-law function for membrane potential values higher than a certain threshold (see materials and methods). one way to change this threshold would be to elevate intrinsic noise which would perhaps result in the shift of this threshold, and thereby, preserving the contrast invariance for a larger range of contrasts.
Mechanisms other than noise that can promote contrast-invariance in our model space with default parametrization is by allowing low-contrast stimuli of preferred orientation to result in spiking responses whereas high-contrast stimuli of the orthogonal orientation not resulting in any spiking activity at all.

\emph{Species difference in orientation selectivity of neurons}

In cat V1, the excitation and inhibition conductances are organized in push-pull model, where the increase of excitation decreases the inhibition, and vice versa \cite{Anderson2000}. Both inhibition and excitation have similar preferred orientation and tuning width, 
which indicate that the cells receive input from neurons with similar orientation tuning \cite{Anderson2000}. In contrast, the V1 cells in mouse V1 receive inhibitory inputs from the inhibitory neurons, 
whose receptive fields have larger size and weaker orientation 
tuning compared to excitatory neurons \cite{Liu2009}. 
Our model space with feature specific connectivity decreases the orientation selectivity of inhibitory neurons and sharpens the orientation selectivity of excitatory neurons. This result is 
consistent with that described by \cite{Liu2011c} as shown using a K-S
test in figure (\ref{fig:FullValidation} A), which suggested that the broadly tuned inhibitory neuron increases the orientation selectivity of membrane potential plotted in figure (\ref{fig:HistogramTuning}). 
Measuring membrane potential fluctuations can also shed light on the origin of strong OS near pinwheels \cite{Marino2005, Schummers2002} in the orientation map of cats and monkeys \cite{Bonhoeffer_Grinvald_1991, Horton_Adams_2005}. If the probability of connection depends mostly on the anatomical distance, neurons near pinwheels will integrate inputs from cells with diverse POs. In our model, when we measure the total aggregate input excitatory plus inhibitory currents to V1 cells we find 
integration of input is comparable to pinwheels in cat. In line with the proposal that each V1 neurons in layer 2/3 integrate spatially distributed inputs which code for many stimulus orientations \cite{Jia_Rochefort2010}. It is important to emphasize that aggregate input 
is determined primarily by two factors: (a) the diversity in local map OSI of the individual presynaptic cells, (b) the tuning width of those presynaptic inputs. In order incorporate diverse tuning width in the model with afferent specificity, we have splitted excitatory afferent input into two groups. Each group is tuned differently than the other group resulting into parallel input channels that carries information about different tuning of orientation to a group of inhibitory neurons located in V1. This model assumption is in line with the recent finding from Gao et.al. (Gao 2011) where they argue neurons in mouse V1 receive inputs from a weighted combination of parallel afferent pathways with distinct spatiotemporal sensitivities. These model assumptions could demonstrate that inhibitory neurons have similar degrees of orientation selectivity as excitatory neurons in V1 consistent with the finding from \cite{Runyan2010}. 

In conclusion, we make the following testable predictions,
(a) Feedforward excitatory inputs to interneurons in mouse V1 are  divided into tuned and untuned channels. 
In general the role of inhibition in generating response selectivity as a
component of the cortical computation remains elusive. With afferent specificity we find shared selectivity for excitation and inhibition 
may act to sharpen orientation selectivity. Inhibition may also
contribute to selectivity by acting as a gain control mechanism.
Our model prediction can be tested by looking at the firing rate response gain for V1 inhibitory cell types and see if there exist any observable difference between the cell types that receives tuned versus untuned afferent presynaptic inputs. 
\section*{Acknowledgments}
The authors are grateful to the CRCNS collaborative research grant and BMBF agency for
funding this work.

\bibliographystyle{CUEDbiblio}
\newpage

\section*{Figures}


\begin{table}[ht!]
\begin{tabular}{lll}
Response properties & Datasets & P-values\\
\hline
\multicolumn{3}{l}{P values based on K-S test}\\
\hline
Excitatory neurons $V_{m}$, $g_{e}$ and $g_{i}$ selectivity\\
$V_{m}$ & Tan vs Liu & 0.0235\\
$g_{e}$ & same &0.6947\\
$g_{i}$ & same &0.1548\\
\hline
Excitatory neurons firing rate $f_{exc}$ selectivity\\
$f_{exc}$&Tan vs Liu & $\le 0.001$\\
$f_{exc}$&Tan vs Runyan&0.0024\\
$f_{exc}$&Tan vs Niell & $\le 0.001$\\
$f_{exc}$&Liu vs Runyan & 0.7274\\
$f_{exc}$&Liu vs Niell&0.0307  \\
$f_{exc}$&Runyan vs Niell & 0.0068\\
\hline
Inhibitory neurons firing rate ${f_{inh}}$ selectivity \\
$f_{inh}$&Runyan vs Niell&0.0497\\
\hline
\end{tabular}
\caption{Summary of computed p values using K-S two sample test between reported orientation selectivity distributions of excitatory and inhibitory population distributions for subthreshold response properties such as membrane potential, input conductances (excitatory and inhibitory) and suprathreshold response such as spikes.}
\label{tab:template}
\end{table}
\begin{table}[ht!]
\begin{tabular}{lll}
Parameter&Description&Value\\
\hline
\multicolumn{3}{l}{Network}\\
$D$&Network dimension&50\\
$N_A$&Afferent to each neuron&20\\
\hline
\multicolumn{3}{l}{Cell Properties}\\
$C_m$&Membrane capacity&0.35nF\\
$g_L^E$&Leak conductance of excitatory cells&15.7nS\\
$g_L^I$&Leak conductance of inhibitory cells&31.4nS\\
$E_L$&Leak reversal potential&-80mV\\
\hline
\multicolumn{3}{l}{Background activity}\\
$g_{bg}^{eo}$&Mean excitatory background conductance&0.56$\cdot g_L$\\
$g_{bg}^{io}$&Mean inhibitory background conductance&1.84$\cdot g_L$\\
$\tau_{bg}^{e}$&Excitatory time constant&2.7ms\\
$\tau_{bg}^{i}$&Inhibitory time constant&10.5ms\\
$\sigma^{e}_{bg}$&Standard deviation of excitatory conductance&0.01$\cdot g_L$\\
$\sigma^{i}_{bg}$&Standard deviation of inhibitory conductance&0.01$\cdot g_L$\\
$E^e_{bg}$&Reversal potential of excitatory conductance&-5mV\\
$E^i_{bg}$&Reversal potential of inhibitory conductance&-70mV\\
\hline
\multicolumn{3}{l}{Connectivity}\\
$N_{EE} = N_{IE}$&Excitatory synaptic connections per cell&100\\
$N_{EI} = N_{II}$&Inhibitory synaptic connections per cell&50\\
$\sigma_E=\sigma_I$&Spread of recurrent connections (std. dev.)&4 pixels (50 $\mu m$)\\
\hline
\multicolumn{3}{l}{Synaptic properties}\\
$E_e$&Reversal potential excitatory synapses&0mV\\
$E_i$&Reversal potential inhibitory synapses&-80mV\\
$\tau_E$&Time constant of AMPA-like synapses&5ms\\
$\tau_I$&Time constant of GABA$_A$-like synapses&5ms\\
$\tau_1$&Time constant of NMDA-like synapses&80ms\\
$\tau_2$&Time constant of NMDA-like synapses&2ms\\
$\mu^{delay}_E$&Mean excitatory synaptic delay&4ms\\
$\sigma^{delay}_E$&Standard deviation of excitatory synaptic delay&2ms\\
$\mu^{delay}_{I}$&Mean inhibitory synaptic delay&1.25ms\\
$\sigma^{delay}_{I}$&Standard deviation of inhibitory synaptic delays&1ms\\
\hline
\multicolumn{3}{l}{Afferent synaptic strengths}\\
$\bar g^{Aff}_E$&Afferent peak conductance to excitatory cells&9$\cdot5\cdot g_L^E$\\
$\bar g^{Aff}_I$&Afferent peak conductance to inhibitory cells&0.73$\cdot  \bar g^{Aff}_E$\\
\hline
\multicolumn{3}{l}{Recurrent synaptic strengths}\\
$\bar g_{II}$&Peak conductance from inh. to inh. cells&1.33$\cdot \bar g^{Aff}_E$\\
$\bar g_{EI}$&Peak conductance from inh. to exc. cells&1.33$\cdot \bar g^{Aff}_E$\\
\hline
\end{tabular}
\caption{Parameters of Hodgkin-Huxley network model}
\label{tab:HHModel}
\end{table}
\begin{table}[ht!]
\begin{tabular}{lcccc}
Gating var. & $c_1$ [mV$^{-1}$] & $c_2(V)$ [mV] & $c_3(V)$ [mV] & $c4$\\ 
\hline
$\alpha_{Na\:act\:1}$ & $0.32$ & $-(V + 45)$ & $-(V + 45)/4$ & $-1$ \\
$\alpha_{Na\:inac\:1}$ & $0.128$ & $1$ & $(V + 51)/18$ &  $0$ \\
$\alpha_{Kd\:act\:1}$ & $0.032$ & $- (V +40)$ & $- (V +40)/5$ &  $-1$ \\
$\alpha_{M\:act\:1}$ & $-0.0001$ & $V + 30$ & $(V - 30)/9$ & $-1$ \\
$\alpha_{Na\:act\:2}$ & $-0.28$ & $- (V + 18)$ & $(V + 18)/9$ & $-1$ \\
$\alpha_{Na\:inact\:2}$ & $4.$ & $1$ & $- (V+28)/5$ & $1$ \\
$\alpha_{Kd\:act\:2}$ & $0.5$ & $1$ & $(V+45)/40 $ & $0$ \\
$\alpha_{M\:act\:2}$ & $0.0001$ & $V+30$ & $(V+30)/9$ & $-1$ \\
\hline
\end{tabular}
\caption{Expressions for channel dynamics}
\label{tab:receptor}
\end{table}


\end{document}